\documentclass[aps,prc,twocolumn,superscriptaddress]{revtex4-2}
\usepackage{CJK}
\usepackage{graphicx}% Include figure files
\usepackage{hyperref}
\usepackage{amssymb}
\usepackage{amsmath}
\usepackage[normalem]{ulem}
\usepackage{url}
\usepackage{color}
\usepackage{multirow}
\usepackage{float}
\usepackage{xspace}
\usepackage{amsthm,amsmath,amssymb}
\usepackage{mathrsfs}
\usepackage{diagbox}

\begin{document}
\begin{CJK*} {UTF8}{} %{GB} {gbsn}
%\preprint{APS/123-QED}

\title{An extended Skyrme momentum dependent potential in asymmetric nuclear matter and transport models}
%\thanks{A footnote to the article title}%
\author{Junping Yang}
\affiliation{China Institute of Atomic Energy, Beijing 102413, China}

%\author{Yongjia Wang}
%\affiliation{School of Science, Huzhou University, Huzhou 313000, China}
\author{Xiang Chen}
\affiliation{China Institute of Atomic Energy, Beijing 102413, China}
%\author{Cheng-Jun Xia}
%\affiliation{School of Information Science and Engineering, Zhejiang University Ningbo Institute of Technology, Ningbo 315100, China}
\author{Ying Cui}
%\email{cuiying@ciae.ac.cn}
\affiliation{China Institute of Atomic Energy, Beijing 102413, China}

\author{Yangyang Liu}
\affiliation{China Institute of Atomic Energy, Beijing 102413, China}

\author{Zhuxia Li}
\affiliation{China Institute of Atomic Energy, Beijing 102413, China}

\author{Yingxun Zhang}
\email{zhyx@ciae.ac.cn}
\affiliation{China Institute of Atomic Energy, Beijing 102413, China}
\affiliation{Guangxi Key Laboratory of Nuclear Physics and Technology, Guangxi Normal University, Guilin, 541004, China}

\date{\today}

\begin{abstract}

Based on an extended Skyrme momentum-dependent interaction (MDI), we derive an isospin asymmetric equation of state, isospin-dependent single-particle potential and the Hamiltonian which can be used in the Boltzmann-Uehling-Uhlenbeck (BUU) model and the quantum molecular dynamics (QMD) model at the beam energy less than 1 GeV/u. As an example of the applications of extended Skyrme MDI, we also present the results obtained with the extended Skyrme momentum-dependent interaction in the improved quantum molecular dynamics model (ImQMD), and the influence of the effective mass splitting on the isospin sensitive observables, i.e., the single and double neutron-to-proton ratios, is discussed again.

\end{abstract}

\maketitle
\end{CJK*}

\section{Introduction}

The isospin asymmetric nuclear equation of state (EOS) plays a crucial role in understanding various properties of neutron stars, including the mass-radius relationship~\cite{Steiner2013APJL,Lattimer2012}, tidal deformability~\cite{GW170817}, neutron-star mergers~\cite{GW170817,Baiotti2019,Margalit2019} and core-collapse of supernovae~\cite{LATTIMER2004,Yasin2020,Steiner2013}. Numerous efforts have been made to constrain the isospin asymmetric nuclear EOS, particularly the symmetry energy at densities below 3$\rho_0$~\cite{Chatziioannou2020,Malik2018,Tan2021,BALiuniverse,CYTsang2023,WJXie2019,YXZhang20,Jeremy2019}. However, our understanding of the dependence of the neutron-star EOS on temperature and constituents such as neutrons, protons and baryons, is limited when relying solely on the properties of neutron stars.

The dependence of the symmetry energy on temperature and constituents can be extracted from heavy ion collisions (HICs)~\cite{Hermann2022PPNP,Agnieszka2023PPNP}. Up to now, some important progresses on the constraints of the density dependence of the symmetry energy via HICs have been obtained~\cite{Tsang09PRL,BALiuniverse,Tsang2012PRC,YXZhang20,wangyj2020,Cozma2018EPA,Huth2022}, but the discrepancy between the constraints from HICs and from the neutron stars was also observed~\cite{YYLiu}. One of the possibilities is the momentum-dependent symmetry potential. The different forms of the momentum-dependent symmetry potential can lead to the same density dependence of symmetry energy, which results in the same properties of neutron stars but different effects on the isospin sensitive observables~\cite{BALi04NPA,Rizzo05PRC,YXZhang14PLB,ZQFeng,JunXu,BALi04PRC}.

The momentum-dependent symmetry potential is calculated from the difference between the single-particle potential of the neutron $V_n$ and the proton $V_p$ over the isospin asymmetry of the system $\delta$, i.e., $V_{sym}=(V_n-V_p)/2\delta$~\cite{Lane62}. In general, the single-particle potential $V_{q=n,p}$ is composed of the momentum-independent potential and the momentum-dependent potential. Especially, the isospin-dependent momentum-dependent potential is not clearly known. Thus, constraining the isospin-dependent momentum-dependent potential becomes one of the important topics in heavy ion collisions, and it heavily rely on the transport models. The strategy of constraining the form of neutron and proton single-particle potential with HICs need to assume the form of isospin-dependent MDI in advance, and then simulate the HICs with different isospin-dependent MDI in the transport model. By comparing the calculations to the data, one can constrain the isospin-dependent MDI indirectly. 

The isospin-dependent momentum-dependent potential used in the transport models can be generally divided into three types. The first one is the square-type~\cite{Rizzo05PRC,YXZhang14PLB,ZhenZhang2018,FYWang2023NST}, %which leads the single-particle potential in symmetric nuclear matter as,
\begin{equation}
    V_q(p)\propto \int d^3p'f_q(\mathbf{r},\mathbf{p}') (\mathbf{p}-\mathbf{p}')^2 %a_q \mathbf{p}^2+\mathbf{b}_q\cdot \mathbf{p}+c_q
\end{equation}
where $f_q(\mathbf{r},\mathbf{p}')$ is the phase space distribution function~\cite{Rizzo05PRC,YXZhang14PLB,ZhenZhang2018,FYWang2023NST}, and had been used to study the effective mass splitting in HICs. This form is suitable for the HICs at the beam energy approximately less than 300 MeV/u~\cite{AichelinPR}, since it violates the optical potential extrapolated from the nucleon-nucleus reaction data. To fix this problem, the logarithm-type and Lorentzian-type momentum-dependent potential were proposed and used in the transport models. The isospin-dependent logarithm-type momentum-dependent single-particle potential is, 
\begin{equation}
\label{mdi-log}
V_q(p)\propto \int d^3p' f_q(\mathbf{r},\mathbf{p}') t_4\left[ln(t_5(\mathbf{p}-\mathbf{p}')^2+1))\right]^2,     
\end{equation}
and was mainly used in QMD models~\cite{Aichelin1987PRL,ZhangYX06,YJWang2014PRC,Liuyy2021PRC,ZQFeng}. The third one is the Lorentzian-type, and its single-particle potential is,
\begin{equation}
\label{mdi-lorentz}
 V_q(p)\propto \int d^3 p'\frac{f_q(\mathbf{r},\mathbf{p}')}{1+(\mathbf{p}-\mathbf{p}')^2/\mu^2}.
\end{equation}
%However, both the Logarithm and Lorentz form of the single-particle potential can not be analytically introduced in the QMD-like models\cite{Aichelin,Isse}. 
Its isospin-dependent form was proposed in Ref.~\cite{DasPRC2003} and was widely used in QMD-type~\cite{CozmaPRC2013,Natsumi,Isse2005,SuJun16PRC} and BUU-type models~\cite{DasPRC2003,ChenLW2005PRL,RizzoNPA}. 

Theoretically, Eqs.(\ref{mdi-log}) and (\ref{mdi-lorentz}) can only be calculated numerically in the framework of the QMD-type models, but it will cost huge CPU time. To use the momentum-dependent interaction (MDI) conveniently, one approximate the results from Eqs.(\ref{mdi-log}) and (\ref{mdi-lorentz}) by reformulating it with $|\mathbf{p}_i-\mathbf{p}_j|$~\cite{Hartnack94}, where the $\mathbf{p}_i$ and $\mathbf{p}_j$ are the average momentum of the $i$th and $j$th nucleons, respectively. The effect of the width of the wave packet was not explicitly involved. Therefore, modeling the isospin-dependent momentum-dependent potential with an appropriate shape over a wide beam energy region and explicitly considering the effect of the width of wave packet in the QMD type models is necessary. % before one can draw any firm conclusion on the \textcolor{red}{constraints of} symmetry energy through HICs. 
An extended Skyrme MDI~\cite{Davesne15,RWang18} can meet this requirement since both the single-particle potential and the Hamiltonian with the Gaussian wave packet can be calculated analytically. Furthermore, it will also be useful in the BUU-type models and a pioneer work has been done in Ref.~\cite{RWang18} for giving the nuclear equation of state and single-particle potential in nuclear matter. 

This paper is organized as follows: In Sec.\ref{sec:formula}, the form of the extended Skyrme MDI, the single-particle potential, the equation of state and the corresponding Hamiltonian used in the ImQMD model are given. In Sec.\ref{sec:results}, we present the numerical results on the equation of state, the symmetry energy and the single-particle potential in nuclear matter obtained with the extended Skyrme interactions. Then, we used the extended Skyrme MDI in the ImQMD and the effects of effective mass splitting on the neutron to proton yield ratios are simply rediscussed. Sec.\ref{sec:summary} is the summary and outlook.

\section{Formulae}
\label{sec:formula}

%The standard Skyrme effective interaction~\cite{chabnate} is written as,
%\begin{equation}
%    \begin{aligned}
%        V(\mathbf{r}_1,\mathbf{r}_2)=& t_0(1+x_0P_\sigma)\delta(\mathbf{r})\\
%        &+\frac{1}{2}t_1(1+x_1P_\sigma)\left[ \mathbf{k'}^2\delta(\mathbf{r})+\delta(\mathbf{r})\mathbf{k}^2\right]\\
%    &+t_2(1+x_2P_\sigma)\mathbf{k'}\cdot \delta(\mathbf{r})\mathbf{k}\\
%    &+\frac{1}{6}t_3(1+x_3P_\sigma)\left[\rho(\mathbf{R})\right]^\sigma \delta(\mathbf{r})\\
%    &+iW_0\mathbf{\sigma}\cdot \left[ \mathbf{k}'\times \delta(\mathbf{r}) \mathbf{k}\right].
%    \end{aligned}
%\end{equation}
%The definitions of $\mathbf{r}=\mathbf{r}_1-\mathbf{r}_2$, \textcolor{blue}{$\mathbf{R}=\frac{1}{2}(\mathbf{r}_1+\mathbf{r}_2)$, and $\mathbf{k}$ denotes the operator $(\mathbf{\nabla_{1}-\nabla_{2}})/2i$ acting on the right; whereas, $\mathbf{k^{'}}$ is the operator $-(\mathbf{\nabla_{1}-\nabla_{2}})/2i$ acting on the left.} %$\mathbf{k}'$ is cc of $\mathbf{k}$ and acting on the left. 
%The second and third terms are the nonlocal term, which correspond to the momentum dependent interaction, which is proportional to the square of the relative momentum between two nucleons. 

In the ImQMD model (-Sky version)~\cite{YXZhang14PLB}, the Skyrme potential energy density without the spin-orbit term is used,
\begin{equation}
\label{eq:edf-imqmd}
    u_\text{sky}=u_\text{loc}+u_\text{md}.
\end{equation}
The local potential energy density is
\begin{eqnarray} \label{eq:edfimqmd}
\begin{aligned}
    u_\text{loc} &= \frac{\alpha}{2}\frac{\rho^2}{\rho_0} +\frac{\beta}{\gamma+1}\frac{\rho^{\gamma+1}}{\rho_0^\gamma}+\frac{g_{sur}}{2\rho_0 }(\nabla \rho)^2 \\
    & \quad +\frac{g_{sur,iso}}{\rho_0}[\nabla(\rho_n-\rho_p)]^2 \\
    & \quad +A_\text{sym}\frac{\rho^2}{\rho_0}\delta^2+B_\text{sym}\frac{\rho^{\gamma+1}}{\rho_0^\gamma}\delta^2.
\end{aligned}
\end{eqnarray}
$\rho=\rho_n+\rho_p$ is the nucleon density and $\delta=(\rho_n-\rho_p)/\rho$ is the isospin asymmetry. The $\alpha$ is the parameter related to the two-body term, $\beta$ and $\gamma$ are related to the three-body term, $g_{sur}$ and $g_{sur,iso}$ are related to the surface terms, $A_\text{sym}$ and $B_\text{sym}$ are the coefficients in the symmetry potential and come from the two- and three-body terms~\cite{ZhangYX06,Zhang20FOP}. The non-local potential energy density is
\begin{eqnarray} \label{eq:mdimqmd}
\begin{aligned}
    u_{md} & = C_0\sum_{ij}\int \text{d}^3p\text{d}^3p' f_i(\mathbf{r},\mathbf{p})f_j(\mathbf{r},\mathbf{p'})(\mathbf{p}-\mathbf{p'})^2 \\
    & \quad + D_0\sum_{ij\in n}\int \text{d}^3 p \text{d}^3p' f_i(\mathbf{r},\mathbf{p}) f_j(\mathbf{r},\mathbf{p'})(\mathbf{p}-\mathbf{p'})^2 \\
    & \quad + D_0\sum_{ij\in p}\int \text{d}^3p \text{d}^3p' f_i(\mathbf{r},\mathbf{p}) f_j(\mathbf{r},\mathbf{p'})(\mathbf{p}-\mathbf{p'})^2.
\end{aligned}
\end{eqnarray}
Here, $f_i(\mathbf{r},\mathbf{p})$ is the phase space density distribution of particle $i$, i.e., $f_i(\mathbf{r},\mathbf{p})=\frac{1}{(\pi\hbar)^3}e^{-\frac{(\mathbf{r}-\mathbf{r}_i)^2}{2\sigma_r^2}-\frac{(\mathbf{p}-\mathbf{p}_i)^2}{2\sigma_p^2}}$. The parameters in Eq.(\ref{eq:edfimqmd}) and Eq.(\ref{eq:mdimqmd}) can be calculated directly from the standard Skyrme interaction~\cite{Zhang20FOP,YXZhang20}, and we named it as standard Skyrme interaction or standard Skyrme-type MDI in the following discussions. %By using the standard Skyrme potential energy density, one can obtain the single-particle potential as $V_q=\delta u/\delta f_q$.
% is the momentum-dependent form of momentum-dependent interaction and its dimension is taken as $[MeV]^2$.  

The treatment of collision term and isospin-dependent Pauli blocking effects are the same as those in Refs.~\cite{ZhangYX06,YXZhangPRC2007}. The isospin-dependent in-medium nucleon-nucleon scattering cross sections in the collision term are assumed to be of the form $\sigma_{nn,pp/np}^*=\left(1-\eta \rho / \rho_0\right) \sigma_{nn,pp/np}^{\text {free }}$, with $\eta$ being a parameter and depends on the beam energy. % The isospin dependent $\sigma_{n n / n p}^{\text {free }}$ is taken from Ref.\cite{Cugnon1996} and the in-medium isospin dependent nucleon-nucleon (NN) differential cross section in free space is adopted from Ref.\cite{Cugnon1996}.
The influence of extended MDI on the $\eta$ is not discussed in this work.

\begin{figure}[htbp]
\includegraphics[width=\linewidth]{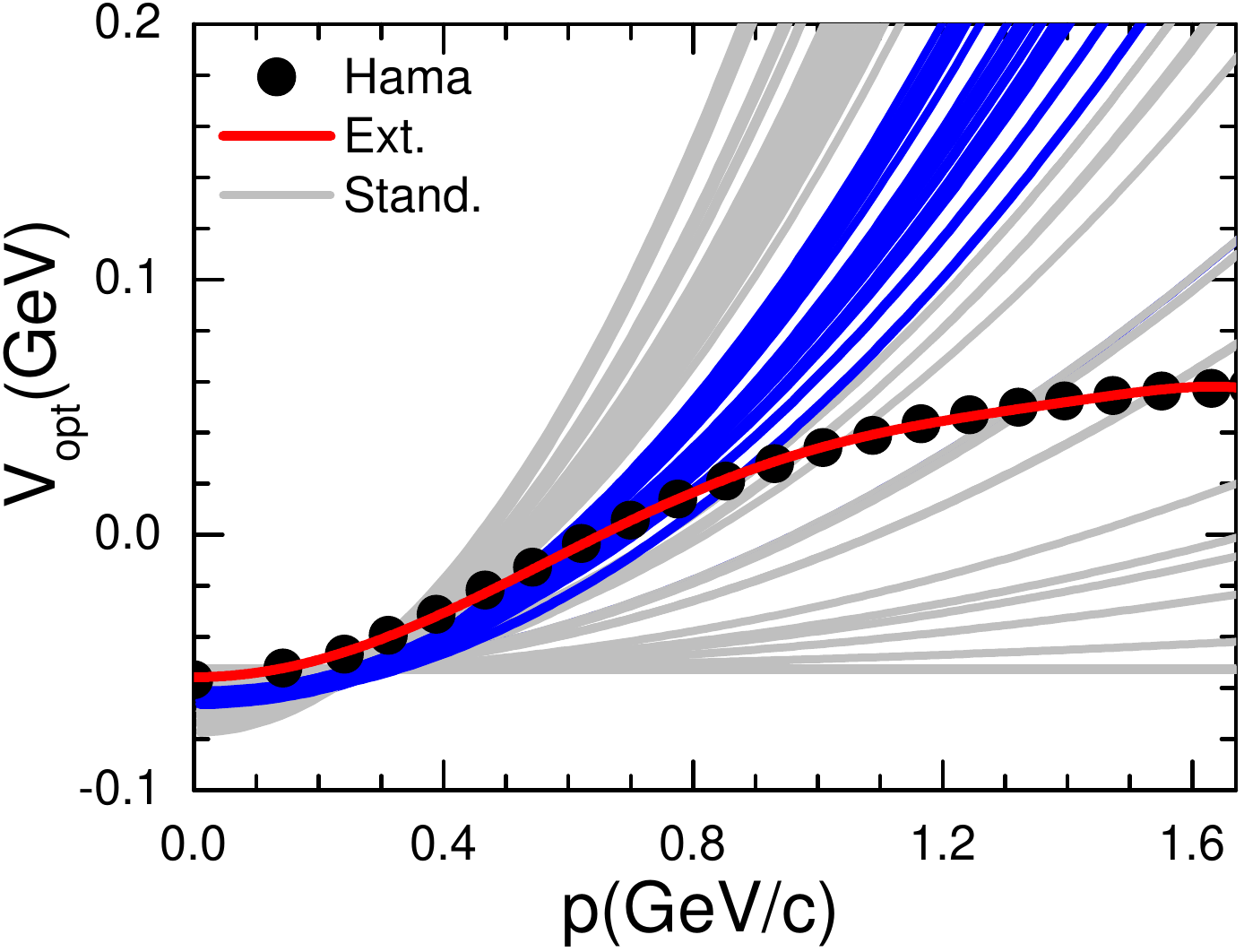}
\caption{single-particle potentials obtained from the standard Skyrme MDI (gray lines and blue lines) and the extended Skyrme MDI (red line). The points are the data of single-particle potential obtained from Hama's data~\cite{hama1990}.}
\label{fig:hama-sky}

\end{figure}

In Fig.\ref{fig:hama-sky}, we present the single-particle potentials in the nuclear matter obtained with the standard Skyrme interactions. The 123 Skyrme parameter sets (gray lines) are selected according to the current knowledge on the nuclear matter parameters~\cite{YXZhang20}, i.e., $K_0=[200,280]$ MeV, $S_0=[25,35]$ MeV, $L=[30,120]$ MeV, $m^*_s/m=[0.6,1.0]$, $f_I=m/m_s^*-m/m_v^*=[-0.5,0.4]$. At the momentum less than 0.8 GeV/$c$ (the kinetic energy is less than 300 MeV), there are 25 Skyrme parameter sets (blue lines): BSk14, BSk15, BSk16, BSk17, BSk9, Gs, MSL0, Rs, Sefm081, SGII, SkM, SKMs, SV-mas08, SkRA, QMC650, QMC700, QMC750, KDE, KDE0v1, SkT7, SkT7a, SkT8, SkT8a, SkT9, SkT9a  can describe the Hama data with reduced $\chi^2$ is less than 0.7. Most of them have $m_n^*>m_p^*$ except for BSk9, KDE, KDE0v1. These single-particle potentials go infinity as the relative momentum increasing, which violates the experimental data of optical potential (solid points)\cite{hama1990}. This condition limits the utility of the standard Skyrme potential energy density in the HICs at the beam energy below 300 MeV/u. Furthermore, the research of the scientific collaboration named UNEDF-SciDAC~\cite{Bertsch, Furnstahl} has shown that there is no more room to improve the standard Skyrme functional to describe the nuclear and neutron stars by simply acting on the optimization procedure.

%The extended Skyrme pseudopotential (NlLO) contains central, spin-orbit, and tensor components. We refer to Refs\cite{Davesne15}. [16,21] for a detailed discussion on the derivation of the(NlLO) pseudopotential. In Cartesian basis it can be written in a familiar form as [22,23]

To overcome the above deficiency of the standard Skyrme interaction, an extended Skyrme effective interaction that includes the terms in relative momenta up to sixth order was proposed in Ref.~\cite{Davesne15}. The extended Skyrme interaction was stimulated by the idea that it is possible to expand a finite range interaction in terms of a zero range like and such an expansion converges~\cite{Carlsson2010}, and the central term can be written order by order as
%\begin{equation}
%    v_\text{Sk}=v_\text{C}+v_\text{T}+v_\text{LS}.
%\end{equation}
%The central term $v_C$ can be written order by order as
\begin{eqnarray}
\label{eq:ext-Sky}
\begin{aligned}
    v_\text{C}&=t_{0}^{(0)}(1+x_{0}^{(0)}P_{\sigma})\delta(\mathbf{r})\\
    &+\frac{1}{2}t_{1}^{(2)}(1+x_{1}^{(2)}P_{\sigma})\delta(\mathbf{r})[\mathbf{k}^{'2}+\mathbf{k}^{2}]\\
    &+t_{2}^{(2)}(1+x_{2}^{(2)}P_{\sigma})\delta(\mathbf{r})\mathbf{k}^{'}\cdot\mathbf{k}\\
    &+\frac{1}{4}t_{1}^{(4)}(1+x_{1}^{(4)}P_{\sigma})\delta(\mathbf{r})[(\mathbf{k}^{'2}+\mathbf{k}^{2})^2+4(\mathbf{k}^{'}\cdot\mathbf{k})^2]\\
    &+t_{2}^{(4)}(1+x_{2}^{(4)}P_{\sigma})\delta(\mathbf{r})(\mathbf{k}^{'}\cdot\mathbf{k})(\mathbf{k}^{'2}+\mathbf{k}^{2})\\
    &+\frac{t_{1}^{(6)}}{2}(1+x_{1}^{(6)}P_{\sigma})\delta(\mathbf{r})(\mathbf{k}^{'2}+\mathbf{k}^{2})\\ 
    &\quad [(\mathbf{k}^{'2}+\mathbf{k}^{2})^2+12(\mathbf{k}^{'}\cdot\mathbf{k})^2]\\
    &+t_{2}^{(6)}(1+x_{2}^{(6)}P_{\sigma})\delta(\mathbf{r})(\mathbf{k}^{'}\cdot\mathbf{k})[3(\mathbf{k}^{'2}+\mathbf{k}^{2})^2+4(\mathbf{k}^{'}\cdot\mathbf{k})^2].
\end{aligned}
\end{eqnarray}
The first term in Eq.(\ref{eq:ext-Sky}) is the local two-body interaction, the rest terms are the non-local interaction with the power of the relative momentum up to six, i.e., $\mathbf{k}^6$. $P_{\sigma}$ is the spin-exchange operator and $\mathbf{k}$ in Eq.(\ref{eq:ext-Sky}) denotes the operator $(\mathbf{\nabla_{1}-\nabla_{2}})/2i$ acting on the right, whereas, $\mathbf{k^{'}}$ is the operator $-(\mathbf{\nabla_{1}-\nabla_{2}})/2i$ acting on the left. They are related to the relative momentum between two nucleons. 

Inspired by this idea, we assume a phenomenological momentum-dependent interaction as
\begin{eqnarray}
\label{eq:ext-mdi}
\begin{aligned}
g(\mathbf{p}-\mathbf{p'})&=\sum_{I=0}^N b_I (\mathbf{p}-\mathbf{p}')^{2I} %b_{0}+b_{1}(\mathbf{p}-\mathbf{p'})^{2}\\\nonumber
%&\quad+b_{2}(\mathbf{p}-\mathbf{p'})^{4}+b_{3}(\mathbf{p}-\mathbf{p'})^{6}+b_{4}(\mathbf{p}-\mathbf{p'})^{8}\\   
%&=\sum_{I=0}^n b_{I}\sum_{k=0}^{2I}\tbinom{2I}{k}\mathbf{p}^k\cdot(-\mathbf{p'})^{2I-k},n=4.\\
\end{aligned}
\end{eqnarray}
in the transport models. The parameter $b_I$ is used to determine the shape of the MDI, and its dimension is GeV$^{2-2I}$ for keeping the dimension of $g(\mathbf{p}-\mathbf{p'})$ in GeV$^2$. Consequently, the energy density $u_{md}$ in Eq.(\ref{eq:mdimqmd}) is replaced by
\begin{eqnarray} \label{eq:extmd-imqmd}
\begin{aligned}
    u_{md} & = \tilde{C}_0\sum_{ij}\int \text{d}^3p\text{d}^3p' f_i(\mathbf{r},\mathbf{p})f_j(\mathbf{r},\mathbf{p'})g(\mathbf{p}-\mathbf{p'}) \\
    & \quad + \tilde{D}_0\sum_{ij\in n}\int \text{d}^3 p \text{d}^3p' f_i(\mathbf{r},\mathbf{p}) f_j(\mathbf{r},\mathbf{p'})g(\mathbf{p}-\mathbf{p'}) \\
    & \quad + \tilde{D}_0\sum_{ij\in p}\int \text{d}^3p \text{d}^3p' f_i(\mathbf{r},\mathbf{p}) f_j(\mathbf{r},\mathbf{p'})g(\mathbf{p}-\mathbf{p'}).
\end{aligned}
\end{eqnarray}
The number of $N$, the interaction parameters $b_I$, $\tilde{C}_{0}$, and $ \tilde{D}_{0}$ in Eq.(\ref{eq:ext-mdi}) and Eq.(\ref{eq:extmd-imqmd}) can be determined by fitting the optical potential data or calculations from microscopic model, such as Dirac-Brueckner-Hartree-Fock~\cite{MutheR1988}. 

At the given shape of MDI, i.e., given $N$ and $b_I$, one can vary the form of symmetry potential, isoscalar effective mass, and the isovector effective mass through $\tilde{C}_{0}$ and $ \tilde{D}_{0}$, which have a dimension in $[fm]^3/[MeV]$. Their values are used to adjust the strength and form of the momentum-dependent symmetry potential for studying the uncertainty of the single-particle potential of neutron and proton in isospin asymmetric nuclear matter.

In the following part, we present the formulas for the single-particle potential and the equation of state, the nuclear matter parameters and their relations to the interaction parameters in transport models and the Hamiltonian used in the QMD-type models as well.

\subsection{single-particle potential and equation of state}
\label{sec:speos}

The single-particle potential $V_q$ in isospin asymmetric nuclear matter can be calculated by taking the functional derivative of the energy density with respect to the single-body phase space distribution of protons or neutrons $f_q(r,p)$,
\begin{equation}\label{sppotential}
\begin{aligned}
 V_q(\rho,\delta,p)&=\frac{\delta u_\text{loc}}{\delta f_q} +\frac{\delta u_\text{md}}{\delta f_q} \\
 &= V^\text{loc}_q(\rho,\delta)+ V^\text{md}_q(\rho,\delta,p),      
\end{aligned}
\end{equation}
where $q$=n or p. The $V_q^{loc}$ is
\begin{equation} \label{vloc}
\begin{aligned}
    V^\text{loc}_q (\rho, \delta) &= \frac{\delta u_\text{loc}}{\delta f_q} \\
    &= \alpha \frac{\rho }{ \rho_0 } + \beta  \frac{ \rho^\gamma }{ \rho_0^\gamma } 
        + (\gamma - 1) B_\text{sym} \frac{\rho^\gamma}{\rho_0^\gamma}\delta^2\\
%        B_\text{sym} (\eta - 1) ( \rho_\text{n} - \rho_\text{p} )^2 \rho^{ 
%        \eta - 2 } \\
%    & \quad \pm 2 A_\text{sym} ( \rho_\text{n} - \rho_\text{p} ) \pm 2 
%        B_\text{sym} ( \rho_\text{n} - \rho_\text{p} ) \rho^{ \eta - 1 }.
    & \quad \pm 2 \left(A_\text{sym} \frac{\rho}{\rho_0}+ 
        B_\text{sym} \left( \frac{\rho}{\rho_0} \right) ^{\gamma} \right) \delta.
\end{aligned}
\end{equation}
The sign `$+$' is for neutrons, and `$-$' for protons. 

The $V_q^{md}$ is,
\begin{eqnarray} \label{vnonlocal}
\begin{aligned}
    V^\text{md}_q(\rho,\delta,p) &= \frac{\delta u_\text{md}}{\delta
        f_q} \\
    &= 2\tilde{C_0}\int d^{3}p^{'}f(\mathbf{r},\mathbf{p'})g(\mathbf{p}-\mathbf{p'})\\
    & \quad +2\tilde{D_0}\int d^{3}p^{'}f_{q}(\mathbf{r},\mathbf{p'})g(\mathbf{p}-\mathbf{p'}).
%    &= 2(C_0+\frac{D_0}{2})\rho\frac{\int d^{3}p^{'}g(\mathbf{p}-\mathbf{p'})}{\int d^{3}p^{'}}
\end{aligned}
\end{eqnarray}
The form of Eqs.(\ref{vloc}) and (\ref{vnonlocal}) can be directly used in transport models for simulating heavy ion collisions by using the phase space density $f_q(\mathbf{r},\mathbf{p})$ obtained in solving the transport equation. Then, the environment of the nucleons changes, such as the density, isospin asymmetry, and temperature of the environment, from the initial to the final stage in the heavy-ion collision will be automatically involved.
%here, $f(\mathbf{r},\mathbf{p'})=f_n(\mathbf{r},\mathbf{p'})+f_p(\mathbf{r},\mathbf{p'})$. 

To quantitatively understand the properties of the extended Skyrme MDI, the single-particle potential and equation of state in the cold nuclear matter are analyzed and compared with that obtained with standard Skyrme interaction. In the cold nuclear matter,
\begin{eqnarray}
%f(\mathbf{r},\mathbf{p'})&=&(\frac{4\pi}{3}p^3_{F})^{-1} \rho \Theta(p_F-|p|)\\\nonumber
%&&=\frac{4}{(2\pi\hbar)^3} \Theta(p_F-|p|),\\\nonumber
f_q(\mathbf{r},\mathbf{p'})&=&(\frac{4\pi}{3}p^3_{F_q})^{-1} \rho_q \Theta(p_{F_q}-|p'|)\\\nonumber
&=&\frac{2}{(2\pi\hbar)^3} \Theta(p_{F_q}-|p'|),    
\end{eqnarray}
and $p_{F_q}=\hbar(3\pi^{2}\rho_q)^{\frac{1}{3}}$. Thus, the corresponding $V^{md}_q$ can also be written in an analytical form:
\begin{equation}\label{vmd-rhoq-series}
    \begin{aligned}
         V^\text{md}_q(\rho,\delta,p) &=2\tilde{C}_0\Bigg[\sum_{I=1}^N b_{I}\sum_{l=0,l\in even}^{2I}\mathcal{\Tilde{A}}_{Il}\sum_{q=n,p}\rho_q^{(2I-l+3)/3} p^l\Bigg]\\
&\quad +2\tilde{D}_0\Bigg[\sum_{I=1}^N b_{I}\sum_{l=0,l\in even}^{2I}\mathcal{\Tilde{A}}_{Il}\rho_q^{(2I-l+3)/3} p^l\Bigg],
    \end{aligned}
\end{equation}
and,
\begin{equation*}
\begin{aligned}
\mathcal{\Tilde{A}}_{Il}&=\frac{2}{(2\pi\hbar)^3}4\pi\Bigg[\left(\hbar(3\pi^2)^{1/3}\right)^{(2I-l+3)} \tbinom{2I}{l}\frac{1}{(2I-l+3)}\Bigg].
\end{aligned}
\end{equation*}
Here, $\tbinom{2I}{l}$ represents the number of combinations of $l$ elements selected from $2I$ elements. One should note that the summation of $b_I$ in Eq.(\ref{vmd-rhoq-series}) starts from $b_1$. The value of $b_0$ will be related to the strength of the local potential when we use the extended Skyrme energy density functional. 

Furthermore, the single-particle potential $V_q$ as in Eq.(\ref{sppotential}) can be expanded as a power series of isospin asymmetry $\delta$, by using the $\rho_q=\frac{\rho}{2}(1+\tau_q\delta)$, with $\tau_q=1$ or $-1$ for neutrons and protons. The so-called Lane potential\cite{Lane62} means one neglects the higher-order terms ($\delta^2$, $\delta^3$, $\cdots$), i.e.,
\begin{equation}\label{eq:vq}
    V_q=V_0\pm V_{sym}\delta+\cdots.
\end{equation}
The $V_0$ is 
\begin{eqnarray}\label{v0}
\begin{aligned}
   V_0&=\alpha \frac{\rho }{ \rho_0 } + \beta  \frac{ \rho^\gamma }{ \rho_0^\gamma }+ 4(\tilde{C}_0+\frac{\tilde{D}_0}{2})\\
   &\quad\times\Bigg[\sum_{I=1}^N b_{I}\sum_{l=0,l\in even}^{2I}\mathcal{\Tilde{A}}_{Il}\times(\frac{\rho}{2})^{(2I-l+3)/3}\times p^l\Bigg].\\
%   V_0&=\alpha \frac{\rho }{ \rho_0 } + \beta  \frac{ \rho^\gamma }{ \rho_0^\gamma } 
%        + (\gamma - 1) B_\text{sym} \frac{\rho^\gamma}{\rho_0^\gamma}\delta^2\\
%   &\quad + 4(\tilde{C}_0+\frac{\tilde{D}_0}{2})\Bigg[\sum_{I=1}^N b_{I}\sum_{l=0,l\in even}^{2I}\mathcal{\Tilde{A}}_{Il}\\
%   &\quad\times(\frac{\rho}{2})^{(2I-l+3)/3}\times p^l\Bigg].\\
%V^\text{md}_0 &=2(\tilde{C}_0+\frac{\tilde{D}_0}{2})\rho\frac{\int_{\mathbf{p'}<p_{F}} d^{3}p^{'}g(\mathbf{p}-\mathbf{p'})} {\int_{\mathbf{p'}<p_{F}} d^{3}p^{'}}\\
%&=2(\tilde{C}_0+\frac{\tilde{D}_0}{2})\left(\frac{4}{(2\pi\hbar)^3}\right)4\pi\\
%&\Bigg[\sum_{I=0}^4 b_{I}\sum_{k=0,k\in even}^{2I} \mathscr{G}_I(k)\rho^{(2I-k+3)/3}\times p^k\Bigg]\\
%\mathscr{G}_I(k)&=\tbinom{2I}{k}\frac{1}{(2I-k+3)}\left(\hbar(\frac{3\pi^2}{2})^{1/3}\right)^{2I-k+3}
\end{aligned}
\end{eqnarray}
%The coefficients $B_i$ are,
%\begin{eqnarray}
%\begin{aligned}
%    B_1&=(2C_0+D_0)b_1\rho+g(8\pi C_0+4\pi D_0)\\
%    &\times\left(\frac{6}{5}A_{const}^5 b_1 \rho^{5/3}+\frac{15}{7}A_{const}^7b_3\rho^{7/3}+\frac{28}{9}b_4A_{const}^9\rho^{3}\right)\\
%    B_2&=(2C_0+D_0)b_2\rho+g(8\pi C_0+4\pi D_0)\\
%    &\times\left(3A_{const}^5 b_2 \rho^{5/3}+10b_3A_{const}^7\rho^{7/3}\right)\\
%    B_3&=(2C_0+D_0)b_3\rho+g(8\pi C_0+4\pi D_0)\frac{28}{5}A_{const}^5 b_3 \rho^{5/3}\\
%    B_4&=(2C_0+D_0)b_4\rho\\    
%    B_0&=(2C_0+D_0)b_0\rho
%\end{aligned}
%\end{eqnarray}
%with $A_{const}=\hbar(\frac{3\pi^2}{2})^{1/3}$
$V_{sym}$ is
\begin{equation}\label{eq:V_mdasy}
    \begin{aligned}
    V_{sym}&=2A_\text{sym} \frac{\rho}{\rho_0}+ 
        2B_\text{sym} \left( \frac{\rho}{\rho_0} \right) ^{\gamma} \\
   &\quad +2\tilde{D}_0\Bigg[\sum_{I=1}^N b_{I}\sum_{l=0,l\in even}^{2I}\mathcal{\Tilde{A}}_{Il} \frac{2I-l+3}{3}\\
&\quad \quad \quad (\frac{\rho}{2})^{(2I-l+3)/3} p^l\Bigg].\\
   \end{aligned}
\end{equation}
The first and second terms in Eq.(\ref{eq:V_mdasy}) are the momentum-independent parts of the symmetry potential, and the last term is the momentum-dependent part of the symmetry potential.
The isospin asymmetric equation of state for cold nuclear matter reads
\begin{equation} \label{eossky}
\begin{aligned}
    E/A(\rho,\delta) &= \frac{3\hbar^2}{10m}\left(\frac{3\pi^2}{2}\rho\right)^{2/3}+\frac{\alpha}{2}\frac{\rho}{\rho_{0}}+\frac{\beta}{\gamma+1}\frac{\rho^{\gamma}} {\rho^{\gamma}_{0}} \\
    &\quad + (\tilde{C}_0+\frac{\tilde{D}_0}{2})\sum_{I=1}^N\tilde{g}_{md}^{(I)}\times\rho^{2I/3+1}\\
    &\quad + S(\rho)\delta^2 +\cdots % \\
%    &= \frac{3\hbar^2}{10m}\left(\frac{3\pi^2}{2}\rho\right)^{2/3}+\frac{\alpha}{2}\frac{\rho}{\rho_{0}}+\frac{\beta}{\gamma+1}\frac{\rho^{\gamma}} {\rho^{\gamma}_{0}} \\
%    &\quad + (\tilde{C}_0+\frac{\tilde{D}_0}{2})
%    \bigg(\tilde{g}_{md}^{(1)}\rho^{5/3}+\tilde{g}_{md}^{(2)}\rho^{7/3}+\cdots\bigg)\\
%    &\quad + S(\rho)\delta^2+\cdots.
\end{aligned}
\end{equation}
The coefficient of $\tilde{g}_{md}^{(I)}$ is
\begin{equation} \label{Emd}
\begin{aligned}
\tilde{g}_{md}^{(I)}&=\frac{4b_{I}}{2^{(2I+6)/3}}\sum_{l=0,l\in even}^{2I}\mathcal{G}_{Il},
\end{aligned}
\end{equation}
with the $\mathcal{G}_{Il}$,
\begin{equation*}
\begin{aligned}
\mathcal{G}_{Il}&=\left(\frac{2}{(2\pi\hbar)^3}\right)^2(4\pi)^2\Bigg[\left(\hbar(3\pi^2)^{1/3}\right)^{(2I+6)}\\
&\quad \tbinom{2I}{l}\frac{1}{(l+3)}\frac{1}{(2I-l+3)}\Bigg],   
\end{aligned}
\end{equation*}
which has a constant value at given $I$ and $l$ and is independent of $b_I$. 
%is the $I^{th}$ coefficient of extended Skyrme MDI term in the symmetric EOS. 

The density dependence of the symmetry energy $S(\rho)$ becomes
\begin{equation} \label{SE-skyrme}
\begin{aligned}
    S(\rho) & = \frac{\hbar^2}{6m}\left(\frac{3\pi^2\rho}{2}\right)^{2/3}+A_\text{sym}\frac{\rho}{\rho_{0}}+B_\text{sym}\left(\frac{\rho}{\rho_{0}}\right)^{\gamma} \\
    &\quad+ \sum_{I=1}^N\tilde{C}_{sym}^{(I)}\rho^{2I/3+1}. %\\
%    &= \frac{\hbar^2}{6m}\left(\frac{3\pi^2\rho}{2}\right)^{2/3}+A_\text{sym}\frac{\rho}{\rho_{0}}+B_\text{sym}\left(\frac{\rho}{\rho_{0}}\right)^{\gamma} \\
%    & + \Bigg(\tilde{C}_{sym}^{1}\rho^{5/3}+\tilde{C}_{sym}^{2}\rho^{7/3}+\tilde{C}_{sym}^{3}\rho^{3}+\cdots \Bigg )
\end{aligned}
\end{equation}
Here, the $\tilde{C}_{sym}^{(I)}$ is taken as
\begin{equation} \label{Smd}
\begin{aligned}
\tilde{C}_{sym}^{(I)}&=2\tilde{C}_0\Bigg[\frac{b_{I}}{2^{(2I+6)/3}}\sum_{l=0,l\in even}^{2I}\mathcal{G}_{Il}\\
&\quad \Big(\frac{l+3}{3}\frac{l}{3}+\frac{2I-l+3}{3}\frac{2I-l}{3}\Big)\Bigg]\\
&\quad +\tilde{D}_0\Bigg[ \frac{b_{I}}{2^{(2I+6)/3}}\frac{2I+6}{3}\frac{2I+3}{3}\sum_{l=0,l\in even}^{2I} \mathcal{G}_{Il}\Bigg].\\
\end{aligned}
\end{equation}
which is the I-th coefficient of the extended Skyrme-type MDI term in the symmetry energy. 
%For more detailed derivation, please refer to the appendix \eqref{ExtMDI}.
%Thus, two coefficients $\tilde{g}_{md}^{I}$ and $\tilde{C}_{sym}^{I}$ are determined by the $b_I$, $\tilde{C}_0$ and $\tilde{D}_0$.
%\sout{The derivations of the single-particle potential and EOS can be found in the Appendix \ref{appendix:EOS-esky}.}
The derivations of the energy density and the single-particle potential can be found in Appendix \ref{appendix:EOS-esky} and Appendix \ref{appendix:Vq}.

The single-particle potential, EOS and symmetry energy at finite temperature can be obtained with Eqs.(\ref{vloc}),(\ref{vnonlocal}), and Eq.(\ref{eq:extmd-imqmd}) by replacing the $f$ with the Fermi distribution at a certain temperature, and solve these equations with the same method as in Ref.~\cite{OuliPLB2011}. Our calculations show that the difference of single-particle potential between T=0 and T=20 MeV is less than 9\% at the momentum less than 500 MeV/$c$, and similar at the momentum greater than 500 MeV/$c$.

\subsection{Nuclear matter parameters and its relation to the interaction parameters}
\label{sec:NM-Intcoeff}

To investigate the uncertainty of the neutron-proton single-particle potential in isospin asymmetric nuclear matter, one may vary it by adjusting the effective nuclear interaction parameters in the transport models. However, directly adjusting the effective nuclear interaction parameters alway mixes the contributions from the isoscalar and isovector part of single-particle potential. To isolate the contributions from the isoscalar and isovector part of the effective interaction, one can set the nuclear matter parameters as an input variable in the transport models. The advantages of using nuclear matter parameters as input are as similar as in Refs.~\cite{WCChen14,YXZhang20,Morfouace2019}. First, the nuclear matter parameters have a precise physical meaning. Second, the Bayesian analysis in transport model simulations in the nuclear matter parameter space can be easily performed, and the uncertainty of the theoretical predictions for all bulk properties can be directly obtained.

In this work, the parameters used in the extended Skyrme interaction as in Eq.(\ref{eossky}), i.e., $\alpha$, $\beta$, $\gamma$, $A_{sym}$, $B_{sym}$, $\tilde{C}_0$, and $\tilde{D}_0$, can be obtained from the nuclear matter parameters by the same standard protocol as in the standard Skyrme interaction\cite{YXZhang20}. They are realized by solving the seven equations for the determination of the saturation density $\rho_0$, the energy per nucleon at saturation density $E_0$, the incompressibility $K_0$, the isoscalar effective mass $m_s^*$, the isovector effective mass $m_v^*$, the symmetry energy coefficient $S(\rho_0)$ and the slope of the symmetry energy $L$.
%The pressure in the nuclear fluid can be calculated as follows:
%\begin{equation}
%    P=\rho^2\frac{\partial E/A(\rho,\delta)}{\partial \rho}.
%\end{equation}

The first equation is related to the value of the saturation density $\rho_0$ which is obtained by seeking the root of the following equation:
\begin{equation}
\begin{aligned}
    P&=\rho_0^2\frac{d}{d\rho}\frac{E}{A}(\rho_0,\delta=0)=0,
\end{aligned}
\end{equation}
i.e.,
\begin{equation}\label{rho0}
    \begin{aligned}
    &\frac{2}{5}\epsilon_{F}^0\rho_0+\frac{\alpha}{2}\rho_0+\frac{\beta}{\gamma+1}\gamma\rho_0\\
    &\quad \quad +(\tilde{C}_0+\frac{\tilde{D}_0}{2})\sum_{I=1}^N\tilde{g}_{md}^{(I)}\left(\frac{2I}{3}+1\right)\rho_0^{2I/3+2}=0,
    \end{aligned}
\end{equation}
here, $\epsilon_{F}^0=\frac{\hbar^2}{2m}\left(\frac{3\pi^2\rho_0}{2}\right)^{2/3}$.

The second equation is the binding energy $E_0$. It reads
\begin{equation}\label{E0}
\begin{aligned}
E_0&=E/A(\rho_0)\\
&=\frac{3}{5}\epsilon_{F}^0+\frac{\alpha}{2}+\frac{\beta}{\gamma+1}+(\tilde{C}_0+\frac{\tilde{D}_0}{2})\sum_{I=1}^N\tilde{g}_{md}^{(I)}\rho_0^{2I/3+1}.\\
\end{aligned}
\end{equation}

The third equation is the incompressibility $K_0$, i.e.,
\begin{equation}\label{K0}
\begin{aligned}
K_0&=9\rho_0^2\frac{\partial^2 E/A}{\partial \rho^2}|_{\rho_0}\\
&=\frac{-6}{5}\epsilon_{F}^0+9\frac{\beta}{\gamma+1}\gamma(\gamma-1)\\
&\quad+6(\tilde{C}_0+\frac{\tilde{D}_0}{2})\sum_{I=1}^N\tilde{g}_{md}^{(I)}\left(\frac{2I}{3}+1\right)I\rho_0^{2I/3+1}.
\end{aligned}
\end{equation}

The fourth and fifth equations are related to the neutron and proton effective mass or the isoscalar and isovector effective mass. The neutron and proton effective mass is obtained from the neutron and proton potential according to
\begin{equation}
\label{effm}
    \frac{m}{m_q^*} =1+\frac{m}{p}\frac{\partial V_q}{\partial p} , \quad q=n,p.   
\end{equation}
The neutron and proton effective mass will be 
\begin{equation} \label{effmq}
\begin{aligned}
   \frac{m}{m_q^*}&=1+2\tilde{C}_0m\Bigg[\sum_{I=1}^N b_{I}\sum_{l=0,l\in even}^{2I}\mathcal{\Tilde{A}}_{Il}\\
&\quad  \sum_{q}\rho_{q}^{(2I-l+3)/3}l\times p^{l-2}\Bigg]\\
&\quad +2\tilde{D}_0m \Bigg[\sum_{I=1}^N b_{I}\sum_{l=0,l\in even}^{2I}\mathcal{\Tilde{A}}_{Il}\\
&\quad \quad \rho_q^{(2I-l+3)/3}l \times p^{l-2}\Bigg].
%1+2\tilde{C}_0m\left(\frac{2}{(2\pi\hbar)^3}\right)4\pi\\
%   &\Bigg[\sum_{I=0}^n b_{I}\sum_{k=0,k\in even}^{2I}\tbinom{2I}{k} \left(\hbar((3\pi^2)^{1/3}\right)^{2I-k+3}\\
%&\frac{1}{(2I-k+3)}(\rho_n^{(2I-k+3)/3}+\rho_p^{(2I-k+3)/3})k\times p^{k-2}\Bigg]\\  
%&+2\tilde{D}_0m\left(\frac{2}{(2\pi\hbar)^3}\right)4\pi \\
%&\Bigg[\sum_{I=0}^n b_{I} \sum_{k=0,k\in even}^{2I}\tbinom{2I}{k} \left(\hbar(3\pi^2)^{1/3}\right)^{2I-k+3} \\
%& \frac{1}{(2I-k+3)}\rho_q^{(2I-k+3)/3}k\times p^{k-2}\Bigg].
\end{aligned}
\end{equation}
Then, we can find the relationship between $\Delta m_{np}^*=(m_n^*-m_p^*)/m$ and $V_{sym}$ as
\begin{equation}\label{eq:dmn-vasy}
    \begin{aligned}
      \Delta m_{np}^*\approx -(\frac{m^*}{m})^2 4m\delta \frac{\partial V_{sym}}{\partial p^2}, 
    \end{aligned}
\end{equation}
The derivation can be found in Appendix\ref{dmnp-vasy}. It means that the strength of the effective mass splitting $\Delta m_{np}^*$ depends on the momentum-dependent part of the symmetry potential when the $m^*/m$ is fixed. %One should note that the effective mass of neutron or proton depends not only on the density but also on the momentum.

The isoscalar effective mass $m_s^*$ can be obtained at $\rho_q=\rho/2$ from Eq. \eqref{effmq}, and the isovector effective mass $m_v^*$ can be obtained at $\rho_q=0$ which represents the neutron(proton) effective mass in pure proton(neutron) matter as in Refs.~\cite{chabnate,ZhenZhang2016}. They are

\begin{equation}
\begin{aligned}
\label{ms*mv*}
   \frac{m}{m_s^*}(\rho,p)&=1+4(\tilde{C}_0+\frac{\tilde{D}_0}{2})m\Bigg[\sum_{I=1}^N b_{I}\sum_{l=0,l\in even}^{2I}\mathcal{\Tilde{A}}_{Il}\\
&\quad (\frac{\rho}{2})^{(2I-l+3)/3}l \times p^{l-2}\Bigg],\\
   \frac{m}{m_v^*}(\rho,p)&=1+4\tilde{C}_0m\Bigg[\sum_{I=1}^N b_{I}\sum_{l=0,l\in even}^{2I}\mathcal{\Tilde{A}}_{Il}\\
&\quad  (\frac{\rho}{2})^{(2I-l+3)/3}l\times p^{l-2}\Bigg].\\
\end{aligned}
\end{equation}
As same as in Ref.~\cite{YXZhang14PLB}, we define a quantity ${f}_I$,
\begin{equation}
\begin{aligned}
\label{fi}
%    &f_I=\frac{1}{2\delta} \left( \frac{m}{m_n^*}-\frac{m}{m_p^*} \right) = \frac{m}{m_s^*}-\frac{m}{m_v^*},\\
{f}_I(\rho,p)&= \frac{m}{m_s^*}-\frac{m}{m_v^*}\\
&=2\tilde{D}_0m\Bigg[\sum_{I=1}^N b_{I}\sum_{l=0,l\in even}^{2I}\mathcal{\Tilde{A}}_{Il}\\
&\quad(\frac{\rho}{2})^{(2I-l+3)/3}l \times p^{l-2}\Bigg],\\
%&=\left( \frac{m}{m_n^*}-\frac{m}{m_p^*} \right)\Bigg[(1+\delta)^{(2I-l+3)/3}-(1-\delta)^{(2I-l+3)/3}\Bigg]^{-1}\\
%   &=\tilde{D}_0m\left(\frac{4}{(2\pi\hbar)^3}\right)4\pi\\
%   &\Bigg[\sum_{I=0}^n b_{I}\sum_{k=0,k\in even}^{2I}\tbinom{2I}{k} \left(\hbar(\frac{3\pi^2}{2})^{1/3}\right)^{2I-k+3} \\
%   &\frac{k}{(2I-k+3)}\rho^{(2I-k+3)/3}\times{p}^{k-2}\Bigg],
\end{aligned}   
\end{equation}
to describe the isospin effective mass splitting, which has the opposite sign with $\Delta m_{np}^*$. In practical transport model calculations, $\Delta m_{np}^*$ depends on the power expansion of isospin asymmetry and can not be used to calculate the $\tilde{D}_0$ accurately. Thus, in the determination of interaction parameters in the transport models, we use the values of $m_s^*/m$ and $f_I$. %Compared with the standard Skyrme-type MDI, the relation $f_I=\frac{1}{2\delta}(\frac{m}{m_n^*}-\frac{m}{m_p^*})$ does not hold again.
%\begin{equation}
%\begin{aligned}
%\label{fi}
%&\left( \frac{m}{m_n^*}-\frac{m}{m_p^*} \right)\\
%&=2\tilde{D}_0m\Bigg[\sum_{I=1}^N b_{I}\sum_{k=0,k\in even}^{2I}\mathcal{\Tilde{A}}_{Ik}(\frac{\rho}{2})^{(2I-k+3)/3}k \times p^{k-2}\\
%&\quad\Bigg((1+\delta)^{(2I-k+3)/3}-(1-\delta)^{(2I-k+3)/3}\Bigg)\Bigg],\\
%&\approx 2\tilde{D}_0m\Bigg[\sum_{I=1}^N b_{I}\sum_{k=0,k\in even}^{2I}\mathcal{\Tilde{A}}_{Ik}(\frac{\rho}{2})^{(2I-k+3)/3}k \times p^{k-2}\\
%&\quad \quad \quad \frac{2}{3} (2I-k+3)\delta \Bigg],\\
%\end{aligned}   
%\end{equation}

The sixth and seventh equations are the symmetry energy coefficient $S_0$,
\begin{equation}\label{S0}
\begin{aligned}
    S_0&=S(\rho_0)\\
    &=\frac{1}{3}\epsilon_{F}^0+A_{sym}+B_{sym}+\sum_{I=1}^N\tilde{C}_{sym}^{(I)}\rho_0^{2I/3+1},\\
\end{aligned}
\end{equation}
and the slope of the symmetry energy $L$ is
\begin{equation}\label{L}
\begin{aligned}
    L&=3\rho_0\frac{\partial S(\rho)}{\partial \rho}|_{\rho_0}\\
    &=\frac{2}{3}\epsilon_{F}^0+3A_{sym}+3B_{sym}\gamma\\
    &\quad +3\sum_{I=1}^N\tilde{C}_{sym}^{(I)}\left(\frac{2I}{3}+1\right)\rho_0^{2I/3+1}.
\end{aligned}
\end{equation}

Given the values $S_0$, $L$, $K_0$, $E_0$, and $\rho_0$, ${f}_I^0={f}_I(\rho_0,p_F)$ and $\frac{m}{m_{s0}^*}=\frac{m}{m_s^*}(\rho_0,p_F)$, 
%$\tilde{f}_I^0(\rho_0,p_F)$, $m_s^*(\rho_0,p_F)$, 
the coefficients $\alpha$, $\beta$, $\gamma$, $A_{sym}$, $B_{sym}$, $\tilde{C}_0$, and $\tilde{D}_0$ can be obtained according to the following formulas:

%$\tilde{D}_0$ is related to $f_I^0$ as,
\begin{equation}\label{eq:D0}
\begin{aligned}
   \tilde{D}_0&={f}_I^0/\Bigg[2m\sum_{I=1}^N b_{I}\sum_{l=0,l\in even}^{2I}\mathcal{\Tilde{A}}_{Il}\\
   &\quad \quad \quad (\frac{\rho_0}{2})^{(2I-l+3)/3}l \times p_F^{l-2}\Bigg].
\end{aligned}    
\end{equation}
%$\tilde{C}_0$ can be obtained from Eq. \eqref{ms*mv*}, given the value of $\frac{m}{m_{s0}^*}$ and $\tilde{D}_0$. It reads,
\begin{equation}\label{eq:C0}
\begin{aligned}
   \tilde{C}_0&=(\frac{m}{m_{s0}^*}-1)/\Bigg[4m\sum_{I=1}^N b_{I}\sum_{l=0,l\in even}^{2I}\\
   &\quad \quad \mathcal{\Tilde{A}}_{Il} (\frac{\rho_0}{2})^{(2I-l+3)/3} l \times p_F^{l-2}\Bigg]-\frac{\tilde{D}_0}{2}.
\end{aligned}   
\end{equation}
%then, we can obtain the terms of $\tilde{g}_{md}^{Ik}$ and $\tilde{C}_{sym}^{Ik}$ form the values of $\tilde{C}_0$ and $\tilde{D}_0$ after we given the number of expansions N of the extended MDI.

%According to the equtions of the saturation density $\rho_0$ (Eq.(\ref{rho0})), energy per nucleon $E/A(\rho_0)$ (Eq.(\ref{E0})), the incompressibility $K_0$ (Eq.(\ref{K0})) at saturation density and the term of $\tilde{g}_{md}^{Ik}$, the parameters of $\alpha$, $\beta$, and $\gamma$ can be obtained by the following relationship,
%Then, we have,
\begin{equation}\label{eq:abg-asymbsym}
    \begin{aligned}
&\gamma=\frac{K_0+\frac{6}{5}\epsilon_{F}^0-6(\tilde{C}_0+\frac{\tilde{D}_0}{2})\sum_{I=1}^N(\frac{2I}{3}+1)I\rho_{0}^{\frac{2I}{3}+1}\tilde{g}_{md}^{(I)}}{\frac{9}{5}\epsilon_{F}^0-6(\tilde{C}_0+\frac{\tilde{D}_0}{2})\sum_{I=0}^NI\rho_{0}^{\frac{2I}{3}+1}\tilde{g}_{md}^{(I)}-9E_0},\\
&\beta=\frac{(\frac{1}{5}\epsilon_{F}^0-\frac{2}{3}(\tilde{C}_0+\frac{\tilde{D}_0}{2})\sum_{I=1}^N I\rho_{0}^{\frac{2I}{3}+1}\tilde{g}_{md}^{(I)}-E_0)(\gamma+1)}{\gamma-1},\\
&\alpha=2E_0-2(\tilde{C}_0+\frac{\tilde{D}_0}{2})\sum_{I=1}^N \rho_{0}^{\frac{2I}{3}+1}\tilde{g}_{md}^{(I)}-\frac{6}{5}\epsilon_{F}^0-\frac{2\beta}{\gamma+1},\\
&B_{sym}=\frac{3S_0-L-\frac{1}{3}\epsilon_{F}^0+2\sum_{I=1}^N I\tilde{C}_{sym}^{(I)}\rho_{0}^{\frac{2I}{3}+1}}{-3(\gamma-1)},\\  
&A_{sym}=S_0-\frac{1}{3}\epsilon_{F}^0-B_{sym}-\sum_{I=1}^N \tilde{C}_{sym}^{(I)}\rho_{0}^{\frac{2I}{3}+1}.
    \end{aligned}
\end{equation}
%According to the equtions of the symmetry energy coefficient $S_0$ (Eq.(\ref{S0})), the slope of the symmetry energy $L$ (Eq.(\ref{L})) at saturation density and the term of $\tilde{C}_{sym}^{Ik}$, the parameters of $A_{sym}$ and $A_{sym}$ can be obtained by the following relationship,
%\begin{equation}
%\begin{aligned}
%&B_{sym}=\frac{3S_0-L-\frac{1}{3}\epsilon_{F}^0+2\sum_{I=1}^N I\tilde{C}_{sym}^{I}\rho_{0}^{\frac{2I}{3}+1}}{-3(\gamma-1)}\\  
%&A_{sym}=S_0-\frac{1}{3}\epsilon_{F}^0-B_{sym}-\sum_{I=1}^N \tilde{C}_{sym}^{I}\rho_{0}^{\frac{2I}{3}+1}
%\end{aligned}
%\end{equation}

%The advantages of using nuclear matter parameters as input are as similar as in Ref.~\cite{WCChen14}. Simply, the nuclear matter parameters have a precise physical meaning. Then, it is easy to do the Bayesian analysis in transport model simulations in the nuclear matter parameter space, and the culmination of the optimization procedure provides theoretical predictions for all bulk properties with meaningful error bars.

%Based on current knowledge of the isospin asymmetric nuclear matter, the nuclear incompressibility $K_0$, the isoscalar effective mass $m_s^*/m$, the isovector effective mass $m_v^*/m$, the symmetry energy coefficient $S_0$, and the slope of symmetry energy $L$ still have certain uncertainties\cite{YXZhang20}. Thus, constraining them in multi-dimensional parameter space is urged. To do it, one has to choose the 

%  \multicolumn{8}{c}{The standard Skyrme-type of MDI:}\\\hline
% &  Para. & $ \alpha $ & $ \beta $ & $ \gamma $ & $ A_{sym}$ &$B_{sym}$ &$\tilde{C}_0$ & $\tilde{D}_0$\\\hline

\subsection{Hamiltonian from extended Skyrme MDI in the quantum molecular dynamics type models}
\label{sec:H-qmd}

Here, we give an example of the Hamiltonian of the extend Skyrme MDI in the QMD-type models. The energy density of the extended Skyrme MDI can be obtained by folding the interaction with the wave function or with the phase space density as in Eq.(\ref{eq:extmd-imqmd}).
%\begin{eqnarray} \label{eq:MDI-QMD}
%\begin{aligned}
%    u_{md} & = \tilde{C}_0\sum_{ij}\int \text{d}^3p\text{d}^3p' f_i(\mathbf{r},\mathbf{p})f_j(\mathbf{r},\mathbf{p'})g(\mathbf{p}-\mathbf{p'}) \\
%    & \quad + \tilde{D}_0\sum_{ij\in n}\int \text{d}^3 p \text{d}^3p' f_i(\mathbf{r},\mathbf{p}) %f_j(\mathbf{r},\mathbf{p'})g(\mathbf{p}-\mathbf{p'}) \\
%    & \quad + \tilde{D}_0\sum_{ij\in p}\int \text{d}^3p \text{d}^3p' f_i(\mathbf{r},\mathbf{p}) f_j(\mathbf{r},\mathbf{p'})g(\mathbf{p}-\mathbf{p'}),
%\end{aligned}
%\end{eqnarray}
%The integration term $I_{ij}=\int \text{d}^3p\text{d}^3p' f_i(\mathbf{r},\mathbf{p})f_j(\mathbf{r},\mathbf{p'})g(\mathbf{p}-\mathbf{p'})$ in Eq.(\ref{eq:extmd-imqmd}) can be obtained analytically. 

In the framework of the ImQMD model, the result of the integration term $I_{ij}=\int \text{d}^3p\text{d}^3p' f_i(\mathbf{r},\mathbf{p})f_j(\mathbf{r},\mathbf{p'})g(\mathbf{p}-\mathbf{p'})$ in Eq.(\ref{eq:extmd-imqmd}) is
\begin{equation}
\label{eq:Iij}
\begin{aligned}
&I_{ij}=\int \text{d}^3p\text{d}^3p' f_i(\mathbf{r},\mathbf{p})f_j(\mathbf{r},\mathbf{p'})g(\mathbf{p}-\mathbf{p'}) \\
&= \frac{1}{(2 \pi\sigma_r^2)^{3}}\frac{1}{(\pi\sigma_p^2)^{1/2}} \exp{[-\frac{(\mathbf{r}-\mathbf{r}_i)^2}{2\sigma_r^2}-\frac{(\mathbf{r}-\mathbf{r}_j)^2}{2\sigma_r^2}]}\\
&\quad \sum_{I=1}^N b_{I}\sum_{l=0,l\in even}^{2I+1} \tbinom{2I+1}{l}(\mathbf{p}_i-\mathbf{p}_j)^{2I-l}W(l,\sigma_p).
\end{aligned}
\end{equation}
Here, $W(l,\sigma_p)$ are the contributions from the width of the wave packet and they are
\begin{equation}
    \begin{aligned}
        W(0,\alpha=1/4\sigma_p^2)&=\frac{\sqrt{\pi}}{2\alpha^{1/2}}=\sqrt{\pi}\sigma_p,\\
        W(1,\alpha=1/4\sigma_p^2)&=\frac{1}{2\alpha}=2\sigma_p^2,\\
        &\cdots,\\
        W(l,\alpha=1/4\sigma_p^2)&=-\frac{\partial}{\partial \alpha} W(l-2,\sigma_p).
    \end{aligned}
\end{equation}
%$\rm{W}(0,\sigma_p)=\frac{\sqrt{\pi}}{2\alpha^{1/2}}=\sqrt{\pi}\sigma_p$ and $\rm{W}(1,\sigma_p)=\frac{1}{2\alpha}$, $\rm{W}(k,\sigma_p)=-\frac{\partial}{\partial \alpha} \rm{W}(k-2,\sigma_p)$ for $k\ge 2$. $\alpha=\frac{1}{4\sigma_p^2}$

The integration of $I_{ij}$ over the coordinate space is
\begin{equation}
\label{eq:H-QMD}
\begin{aligned}
&\int I_{ij} d^3r=\\ %\int \text{d}^3p\text{d}^3p'\text{d}^3r f_i(\mathbf{r},\mathbf{p})f_j(\mathbf{r},\mathbf{p'})g(\mathbf{p}-\mathbf{p'}) \\
&\quad \rho_{ij} \sum_{I=1}^N b_{I}\sum_{l=0,l\in even}^{2I+1} \tbinom{2I+1}{l}(\mathbf{p}_i-\mathbf{p}_j)^{2I-l}\frac{{W}(l,\sigma_p)}{(\pi\sigma_p^2)^{1/2}},
\end{aligned}
\end{equation}
% here,
% \begin{equation}
% \begin{aligned}
% &\rho_{ij}=\frac{1}{(4\pi\sigma_r^2)^{3/2}}\text{exp}[-\frac{(\mathbf{r_i}-\mathbf{r_j})^2}{4\sigma_r^2}].
% \end{aligned}
% \end{equation}
and thus the corresponding part of Hamiltonian is
\begin{eqnarray} \label{eq:extmd-Hamiltonian}
\begin{aligned}
    H & =2\tilde{C}_0\sum_{ij}\rho_{ij} \sum_{I=1}^N b_{I}\sum_{l=0,l\in even}^{2I+1} \tbinom{2I+1}{l}(\mathbf{p}_i-\mathbf{p}_j)^{2I-l}\\
    &\quad\quad \times\frac{{W}(l,\sigma_p)}{(\pi\sigma_p^2)^{1/2}}\\
    &\quad+(\tilde{C}_0+\tilde{D}_0)\sum_{ij\in q}\rho_{ij}\sum_{I=1}^N b_{I}\sum_{l=0,l\in even}^{2I+1}\tbinom{2I+1}{l}\\
    &\quad\quad \times (\mathbf{p}_i-\mathbf{p}_j)^{2I-l}\frac{{W}(l,\sigma_p)}{(\pi\sigma_p^2)^{1/2}},\\
%    &+\tilde{D}_0\sum_{ij\in p}\rho_{ij}\sum_{I=1}^N b_{I}\sum_{l=0,l\in even}^{2I+1} \tbinom{2I+1}{l}(\mathbf{p}_i-\mathbf{p}_j)^{2I-l}\frac{{W}(l,\sigma_p)}{(\pi\sigma_p^2)^{1/2}},\\
%    & \quad + \tilde{D}_0\sum_{ij\in p}\rho_{ij} \sum_{I=1}^N b_{I}\sum_{l=0,l\in even}^{2I+1} \tbinom{2I+1}{l}(\mathbf{p}_i-\mathbf{p}_j)^{2I-l}\frac{{W}(l,\sigma_p)}{(\pi\sigma_p^2)^{1/2}}.\\
\end{aligned}
\end{eqnarray}

where
\begin{equation}
\begin{aligned}
&\rho_{ij}=\frac{1}{(4\pi\sigma_r^2)^{3/2}}\text{exp}[-\frac{(\mathbf{r_i}-\mathbf{r_j})^2}{4\sigma_r^2}].
\end{aligned}
\end{equation}
%By inserting the integration of Eq.(\ref{eq:Iij}) into the integration of the Eq.(\ref{eq:MDI-QMD}) over the space, the Hamiltonian used in the ImQMD model can be obtained. 

%Where,
%\begin{equation}
%\begin{aligned}
%&f_i(\mathbf{r},\mathbf{p})=\frac{1}{(\pi\hbar)^3}exp[-\frac{(\mathbf{r}-\mathbf{r_i})^2}{2\sigma_r^2}-\frac{(\mathbf{p}-\mathbf{p_i})^2}{2\sigma_r^2}]
%\end{aligned}
%\end{equation}

%The corresponding values of each nuclear matter parameter are listed in the table \eqref{NMpara}.

\section{Results and discussions}
\label{sec:results}

\subsection{Determination of the expansion number \texorpdfstring{$N$}{} and coefficient \texorpdfstring{$b_I$}{}}
%The values of $b_I$ (with I=1 to N) determine the shape of the MDI. The values of $\tilde{C}_0$ and $\tilde{D}_0$ are related to the isoscalar effective mass and isovector effective mass splitting according to Eq.(\ref{eq:D0})-Eq.(\ref{eq:C0}). In details, $\tilde{C}_0$ can zoom in or out the shape of MDI, $\tilde{D}_0$ determines the strength of the effective mass splitting. In the case of fitting Hama's optical potential data for symmetric matter, $\tilde{D}_0=0$. 

The values of $N$, $b_I$, $\tilde{C}_0$, and $\tilde{D}_0$ are obtained by fitting the $V_0$($\rho_0$, $p$) in Eq.(\ref{v0}) to Hama's optical potential data~\cite{hama1990}, which gives the isoscalar effective mass $m_s^*/m=0.77$. One should note that the data can be well described when N$\ge$4. As an example, we present the fitting results with extended Skyrme MDI in Fig.~\ref{fig:hama-sky} as the red line. The values of $b_I$, $\tilde{C}_0$, and $\tilde{D}_0$ for $N=4$, 5 and 6 are listed in Table\ref{Vqpara}. Our calculations show that the results of EOS, single-particle potential, and HIC observables, are independent of the $N$ we used, since they are used to fit the same data. In the following calculations, we will keep $N=4$ and the values of $b_1$ to $b_4$ are fixed. Nevertheless, the extrapolated strength of the single-particle potential above 1 GeV is different for different $N$, and which should be investigated by using the HICs at high beam energies\cite{Nara20} and will not be discussed in this paper.% and the corresponding reduced $\chi^2$ of the fitting is 1.81$\times10^{-3}$. The obtained values of $b_I$, $\tilde{C}_0$ and $\tilde{D}_0$ are listed in Table\ref{Vqpara}. 

\begin{table}[htbp]
\centering
\caption{The parameters of $b_I$ used in extended Skyrme MDI are in GeV$^{2-2I}$ and $\tilde{C}_0$, $\tilde{D}_0$ are $\rm{fm}^{3}\rm{GeV}^{-1}$.}
\label{Vqpara}
\begin{tabular}{cccccccccc} \\ 
\hline
\hline
N & $ b_0 $ & $ b_1 $ & $ b_2 $ & $ b_3 $ & $ b_4 $ & $ b_5 $ & $ b_6 $ &$ \tilde{C}_0 $ & $ \tilde{D}_0 $ \\ \hline
4 & -1.105 & 3.649 & -2.608 & 0.826 & -0.093 & - & - & 0.182 & 0.00 \\ 
\hline
5 & -1.135 & 4.137 & -3.821 & 1.837 & -0.428 & 0.038 & - & 0.182 & 0.00 \\ 
\hline
6 & -1.149 & 4.434 & -4.828 & 3.048 & -1.070 & 0.193 & -0.014 & 0.182 & 0.00 \\ 
%6 & -2.793 & 10.775 & -11.734 & 7.406 & -2.600 & 0.468 & -0.034 & 0.075 & 0.00 \\ 
\hline
\hline
%& 1.456 &-4.727 & 3.031 & -0.437 & 0.123 & -0.220 & 0.164 \\
%Para.2~ & 0.045 & -0.146 & 0.094 & -0.014 & 0.004 & -1.800 & -5.319 \\
\end{tabular}%
\end{table}

%In the following calculations, we will keep the values of $b_1$ to $b_4$ fixed. The values of $\tilde{C}_0$ and $\tilde{D}_0$ will be changed according to the different input values of $m_s^*/m$ and $f_I$. %the value of $b_0$ is related to the standard Skyrme interaction parameters, corresponding to the sum of $\alpha$ term and $\beta$ term in Eq.(\ref{v0}).

%\textcolor{blue}{As shown in Fig. \ref{hamadata}, the red solid line is the result of fitting Hama's optical potential data and the gray lines are the results of 123 sets of standard Skyrme parameters. The selection criteria for 123 sets of standard Skyrme parameters are based on the current knowledge of the nuclear matter parameters ~\cite{YXZhang20}, i.e.,
%\begin{equation}
%\label{MP-Criteria}
%\begin{aligned}
%    200 \text{ MeV} \leqslant K_0 & \leqslant 280 \text{ MeV}, \\
%     25 \text{ MeV} \leqslant S_0 & \leqslant  35 \text{ MeV}, \\
%     30 \text{ MeV} \leqslant  L & \leqslant 120 \text{ MeV}, \\
%    0.6             \leqslant m_s^*/&m \leqslant 1.0,       \\
%   -0.5             \leqslant f_I & \leqslant 0.4.
%\end{aligned}
%\end{equation}
%Compared with the results of optical potential given by standard parameters, the optical potential given by fitting Hama's data is relatively flat in the high momentum region.}

\subsection{single-particle potential, equation of state and symmetry energy}

Now, let us check the neutron and proton single-particle potentials, and the isospin asymmetric nuclear equation of state obtained with the extended Skyrme MDI. For comparisons, we also plot the corresponding results obtained with the standard Skyrme MDI. %The single-particle potential is obtained in the isospin asymmetric nuclear matter with isospin asymmetry $\delta=0.2$, and the equation of state is for the symmetric nuclear matter.
In the calculations, we fix the values of $K_0=230$ MeV, $m_s^*/m=0.77$, $S_0=32$ MeV and vary the $L$ and the $f_I$. In Table~\ref{tab:nmpara-QMD}, we list the corresponding interaction parameters, such as $\alpha$, $\beta$, $\gamma$, $A_{sym}$, $B_{sym}$, $\tilde{C}_0$, and $\tilde{D}_0$, that will be used in the transport models. The values in brackets from the second to the eighth rows represent the values obtained with the standard Skyrme MDI as in Ref.~\cite{YXZhang20}.

\begin{table*}[htbp]
\centering
\caption{The parameters used in the calculations corresponding to $K_0=230$ MeV, $m_s^*/m=0.77$, $S_0=32$ MeV, and different values of $L$ and $f_I$. The parameters $\alpha$, $\beta$, $A_{sym}$, $B_{sym}$ are in MeV. $\tilde{C}_0$ and $\tilde{D}_0$ are $\rm {fm}^{3}\rm{GeV}^{-1}$.}
\label{tab:nmpara-QMD}
\begin{tabular}{cccccc} \\ 
 %\multicolumn{8}{l}{The extended Skyrme-type of MDI:}\\\hline
 \hline
 \hline
 Para.& (L=46, $f_I$=0.3) & (L=46, $f_I$=-0.3) & (L=100, $f_I$=0.3) & (L=100, $f_I$=-0.3)  \\ 
 \hline
 $\alpha$ &\multicolumn{4}{c}{-236.58 (-265.78)}&   \\ 
 $\beta$ &\multicolumn{4}{c}{163.95 (194.93)}&    \\ 
 $\gamma$ &\multicolumn{4}{c}{1.26 (1.22)}& \\
 $A_{sym}$ & 83.65 (108.44) &58.57 (62.73)  &14.41 (25.32) & -10.67 (-20.40) \\ 
 $B_{sym}$ & -79.48 (-103.69) &-30.52 (-35.38) &-10.25 (-20.34) &38.72 (47.96)   \\ 
 $\tilde{C}_0$ & $-7.92\times10^{-4}$ ($-2.08\times10^{-3}$) &0.37 (1.00)  &$-7.92\times10^{-4}$ ($-2.08\times10^{-3}$) & 0.37 (1.00)  \\ 
 $\tilde{D}_0$ & 0.37 (1.00) &-0.37 (-1.00) &0.37 (1.00)  & -0.37 (-1.00) \\ \hline
%  \multicolumn{8}{l}{The standard Skyrme-type of MDI:}\\\hline
% $f_I$=+0.3 & ~-265.78 & ~194.93 & ~1.22 & ~108.44\\ 
% $f_I$=-0.3 & ~-265.78 & ~194.93 & ~1.22 & ~62.73\\ \hline
\end{tabular}%
\end{table*}

%\textcolor{blue}{Therefore, we can arbitrarily select a set of nuclear matter parameters and obtain the corresponding values of parameters $\alpha$, $\beta$, $\gamma$, $ A_{sym}$, $B_{sym}$, $\tilde{C}_0$ and $\tilde{D}_0$ in the ImQMD model through the Eq.\eqref{D0}-Eq.\eqref{NM-ImQMD}.}

%\textcolor{blue}{In this work, We are concerned about the impact of different effective mass splitting on isospin sensitive observations produced by heavy ion collisions. Therefore, we selected two different sets of effective mass splitting $f_I=\pm0.3$ and take fixed values for other nuclear matter parameters, as shown in the table \ref{NMpara}. The corresponding values of parameters used in ImQMD model are shown in the table\ref{NMpara-QMD}.}

%\begin{table}[htbp]
%\centering
%\caption{Two sets of nuclear matter parameters used in this work. The parameters $E_0$, $K_0$, $S_0$, $L$ are in MeV, $\rho_0$ is %fm$^{-3}$. $\frac{m_s^*}{m}$, $\frac{m_v^*}{m}$ and $f_I$ are dimensionless.}
%\label{NMpara}
%\begin{tabular}{ccccccccccc} \\ \hline
% & $ \rho_0 $ & $ E_0 $ & $ K_0 $ & $ S_0 $ & $ L$ &$\frac{m_s^*}{m}$& $\frac{m_v^*}{m}$ & $f_I$&  \\\hline
% & ~0.16~ & ~-16~ & ~230~ & ~32~ & ~46~& ~0.77~&~1.00~&~0.3~  \\ 
% & ~0.16~ & ~-16~ & ~230~ & ~32~ & ~46~& ~0.77~&~0.63~&~-0.3~ \\ \hline
%\end{tabular}%
%\end{table}

\begin{figure}[htbp]
\flushleft
\includegraphics[width=\linewidth]{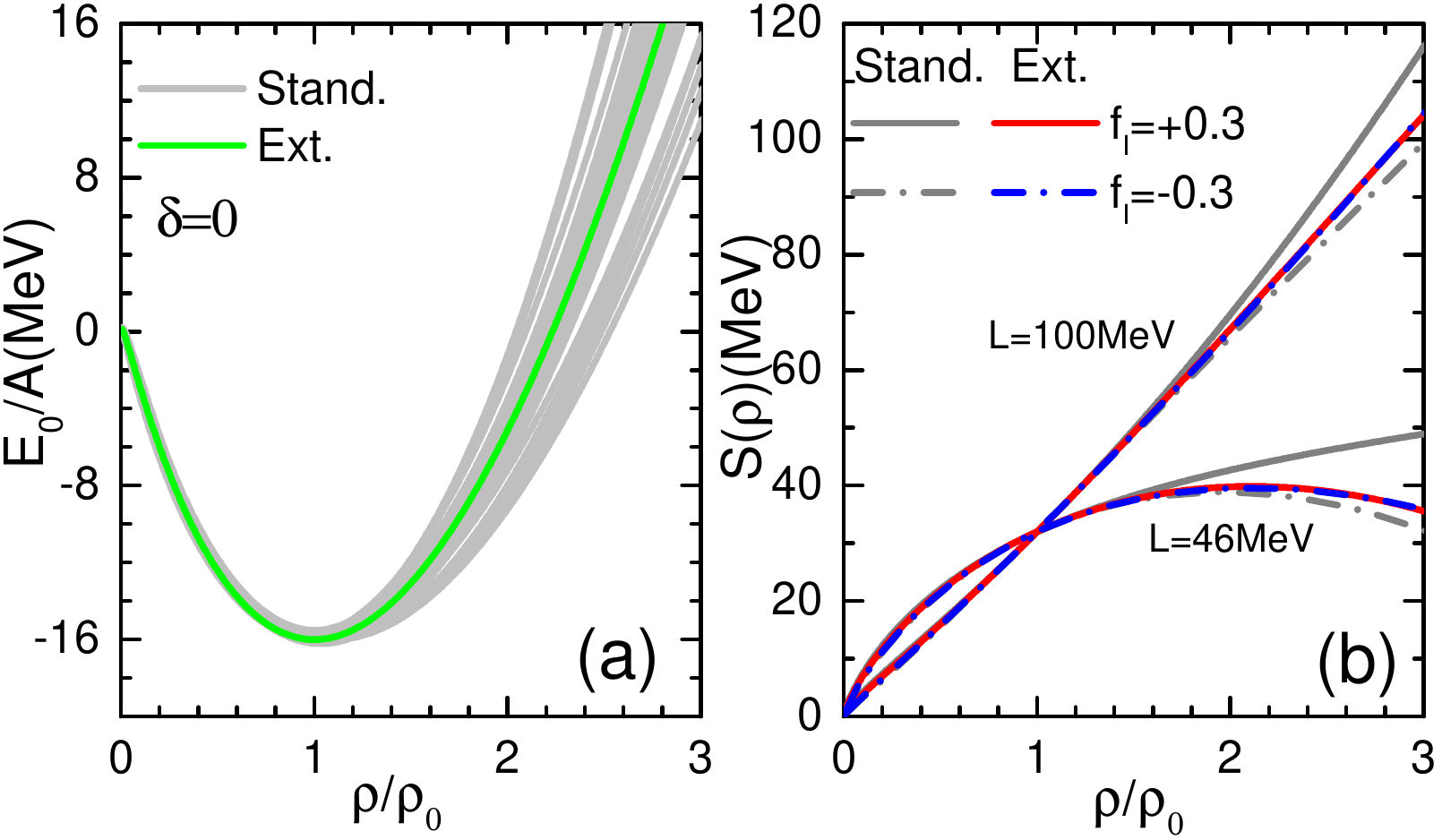}
\setlength{\abovecaptionskip}{0pt}
\vspace{0.2cm}
\caption{(Color online) (a) The equation of state obtained with the extended Skyrme MDI and the standard Skyrme MDI; (b) the density dependence of the symmetry energy obtained with the extended Skyrme MDI and the standard Skyrme MDI, at $f_I=\pm 0.3$ and $L=46, 100$ MeV.}
\setlength{\belowcaptionskip}{0pt}
\label{eos}
\end{figure}

Since the isospin asymmetric nuclear equation of state can be written as,
\begin{equation}
    E(\rho,\delta)/A=E_0(\rho,\delta=0)/A+S(\rho)\delta^2+\cdots
\end{equation}
with a parabolic approximation, the isospin symmetric part of EOS $E(\rho,\delta=0)/A$ is presented in Figure~\ref{eos} (a) and the density dependence of symmetry energy $S(\rho)$ is presented in panel (b). The gray lines in Figure~\ref{eos} (a) are the EOS obtained by the 123 standard Skyrme interaction sets. It shows that the extended Skyrme interaction can reasonably reproduce the EOS for symmetric nuclear matter, and avoid the defect of describing EOS by only using the Taylor expansion parameters as described in Ref.\cite{Margueron18}.

The panel (b) is the density dependence of the symmetry energy obtained with the extended Skyrme MDI (coloured lines) and the standard Skyrme MDI (gray lines). As expected, the results obtained with different $L$ exhibit different density dependence of the symmetry energy. However, the influence of different $f_I$ on the density dependence of the symmetry energy is weak. For the standard Skyrme MDI (gray lines), the influence of different $f_I$ mainly appears at the density above 1.5$\rho_0$. At $\rho=2\rho_0$, the difference between the symmetry energy obtained with two $f_I$ is less than 7\% for $L=100$ MeV, and is less than 13\% for $L=46$ MeV. %The reason is that the small difference on $f_I$ leads to the small difference on $C_{sym}$ as described in Ref.\cite{YXZhang20}. %, thus, the obvious influence of $f_I$ on the density dependence of the symmetry energy appear at large density region. 
For extended Skyrme MDI, the density dependence of the symmetry energy obtained with $f_I=0.3$ (solid lines) and $f_I=-0.3$ (dashed lines) are close to each other at both $L=46$ or 100 MeV at $\rho<3\rho_0$. The reason is that the different effective mass splitting obtained by the extended Skyrme MDI have a smaller impact on the $A_{sym}$, $B_{sym}$, and $C_{sym}^{(I)}$ terms in the density dependent of the symmetry energy compared with the standard Skyrme MDI according to the Eq.(\ref{eq:D0})-Eq.(\ref{eq:abg-asymbsym}). 

Based on the above discussions, one can expect the symmetry energy constraints from the properties of the neutron stars, such as the mass-radius relationship and the tidal deformability, can not well distinguish $f_I$ (or effective mass splitting $\Delta m_{np}^*$), because the Tolman-Oppenheimer-Volkov (TOV) equation of neutron stars only depends on the pressure vs density~\cite{Glendenning}. However, the similar density-dependent symmetry energy from different $f_I$ could lead to different effects on the HICs observables via the momentum-dependent symmetry potential, and thus the constraints of the symmetry energy from HICs may be different from the constraints from neutron stars. %This discrepancy comes from the momentum-dependent symmetry potential, which may degenerate to the same density dependence of symmetry energy, but have different effects on the isospin sensitive HICs observables.

\begin{figure}[htbp]
\flushleft
\includegraphics[width=\linewidth]{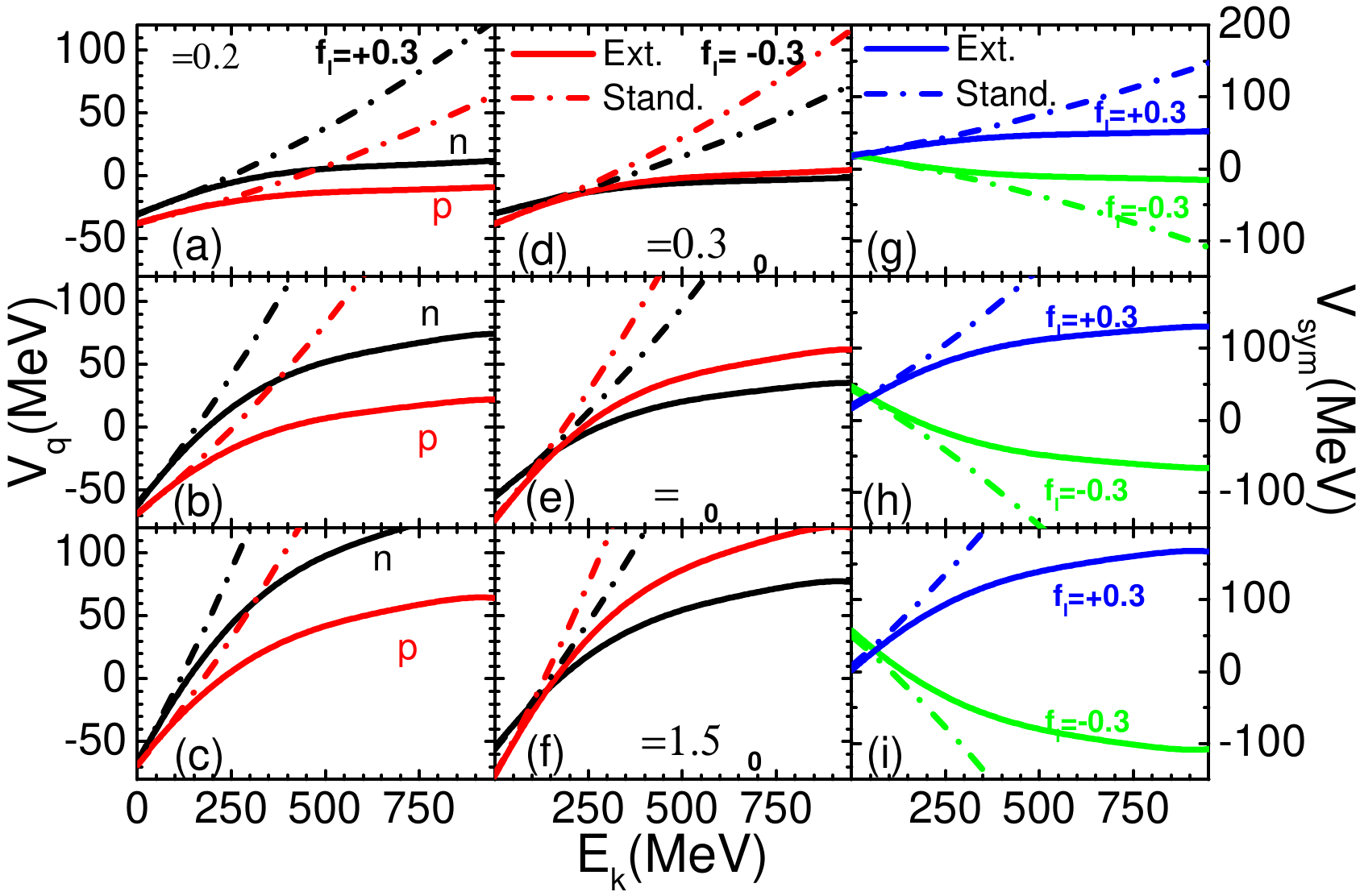}
\setlength{\abovecaptionskip}{0pt}
\vspace{0.2cm}
\caption{(Color online) Left panels and middle panels are the single-particle potential for $f_I=0.3$ and $f_I=-0.3$, respectively. Right panels are the symmetry potential. The upper, middle, and bottom panels are the results at $\rho=0.3\rho_0$, $\rho=\rho_0$\ and $\rho=1.5\rho_0$. Solid lines are the results obtained with extended MDI, and dashed lines are for standard Skyrme MDI.}
\setlength{\belowcaptionskip}{0pt}
\label{fig:VLane}
\end{figure}

To see the neutron and proton single-particle potentials ($V_n$ and $V_p$) in isospin asymmetric nuclear matter obtained with different $f_I$ (or effective mass splitting), we present the $V_n$ and $V_p$ as functions of kinetic energies in nuclear matter with $\delta=0.2$ in Figure~\ref{fig:VLane}. The upper panels, middle panels, and bottom panels are the results obtained at the densities $\rho=0.3\rho_0$, $\rho_0$, and $1.5\rho_0$, respectively. Panels (a)-(c) present the neutron potential $V_n$ (black lines) with $f_I=+0.3$, and panels (d)-(f) are the proton potential $V_p$ (red lines) with $f_I=-0.3$. All these results are obtained according to the Eqs.(\ref{eq:vq})-(\ref{eq:V_mdasy}). The dashed lines are obtained with the standard Skyrme MDI, and the solid lines are obtained with the extended Skyrme MDI. 
\setlength{\abovedisplayskip}{3pt}

As expected, the values of $V_n$ and $V_p$ obtained from the extended Skyrme MDI reach an asymptotic value with an increase of $E_{k}$ at densities we presented. Generally, the strength of single-particle potential increases with the density increasing, and similar trends are for the strength of symmetry potential. There is a cross point of symmetry potential obtained with $f_I=0.3$ and $f_I=-0.3$, and this point moves to a high kinetic energy region with the density increasing. This behavior can be observed by the energy spectral of neutorn to proton yield ratios, i.e., $R_{n/p}$, as discussed in our previous published results~\cite{YXZhang14PLB}.

Focus on the impact of $f_I$, one can observe that the $V_n$ is greater than $V_p$ when $f_I=0.3$ ($m_n^*<m_p^*$). More details, neutrons will feel a stronger repulsive force than protons at high kinetic energy where $V_q>0$, and neutrons feel a weaker attractive force than protons at low kinetic energy where $V_q<0$. For $f_I=-0.3$ ($m_n^*>m_p^*$), $V_n$ is less than $V_p$ at $E_k>150$ MeV and a contradictory behavior is observed at $E_k<150$ MeV. To single out the contributions from the isoscalar single-particle potential, the symmetry potentials $V_{sym}$ as functions of kinetic energy $E_k$ are plotted in Figs.\ref{fig:VLane}(g)-(i). The convention of the line styles is as same as in Figs.\ref{fig:VLane}(a)-(f). The green and blue lines represent $f_I=-0.3$ and $f_I=+0.3$, respectively. The $V_{sym}$ increases (decreases) with the kinetic energy increasing for $f_I=+0.3$ ($f_I=-0.3$). Different than the standard Skyrme MDI, the values of $V_{sym}$ obtained from the extended Skyrme MDI tend to flatten out as $E_k$ increases. Thus, one may expect the difference of the neutron to proton yield ratios in HICs, i.e., $Y(n)/Y(p)$, obtained from two different $f_I$ will become smaller for the extended Skyrme MDI than that for standard Skyrme interaction.

%\begin{figure}[htbp]
%\flushleft
%\includegraphics[width=\linewidth]{Srho.pdf}
%\setlength{\abovecaptionskip}{0pt}
%\vspace{0.2cm}
%\caption{(Color online) Different parameter sets of extended Skyrme-type momentum-dependent interaction were selected to fit the data of Hama et al~\cite{hama1990}.}
%\setlength{\belowcaptionskip}{0pt}
%\label{Srho}
%\end{figure}
%\setlength{\abovedisplayskip}{3pt}

\subsection{The effects of extended Skyrme MDI on the neutron to proton ratios }
\label{sec:imqmd-results}

To see the influence by using the extended Skyrme MDI and the standard Skyrme MDI on the HICs observables, the central collisions of the systems $A=^{124}$Sn+$^{124}$Sn and $B=^{112}$Sn+$^{112}$Sn are simulated at the beam energy of 120 MeV/u and $b=2fm$. The HIC observables, i.e., the single ratio of the coalescence invariant (CI) neutron and proton $R_{n/p}=Y_{CI}(n)/Y_{CI}(p)$ and the double ratio of the coalescence invariant (CI) neutron and proton $DR(n/p)=R_{n/p}(A)/R_{n/p}(B)$, are analyzed under different $L$ and $f_I$. The $Y_{CI}(n)$ and $Y_{CI}(p)$ are obtained by combining the free nucleons with those bound in light isotopes with $1< A <5$~\cite{coupland2016,Morfouace2019}.
\begin{figure}[hbpt]
\flushleft
\includegraphics[width=\linewidth]{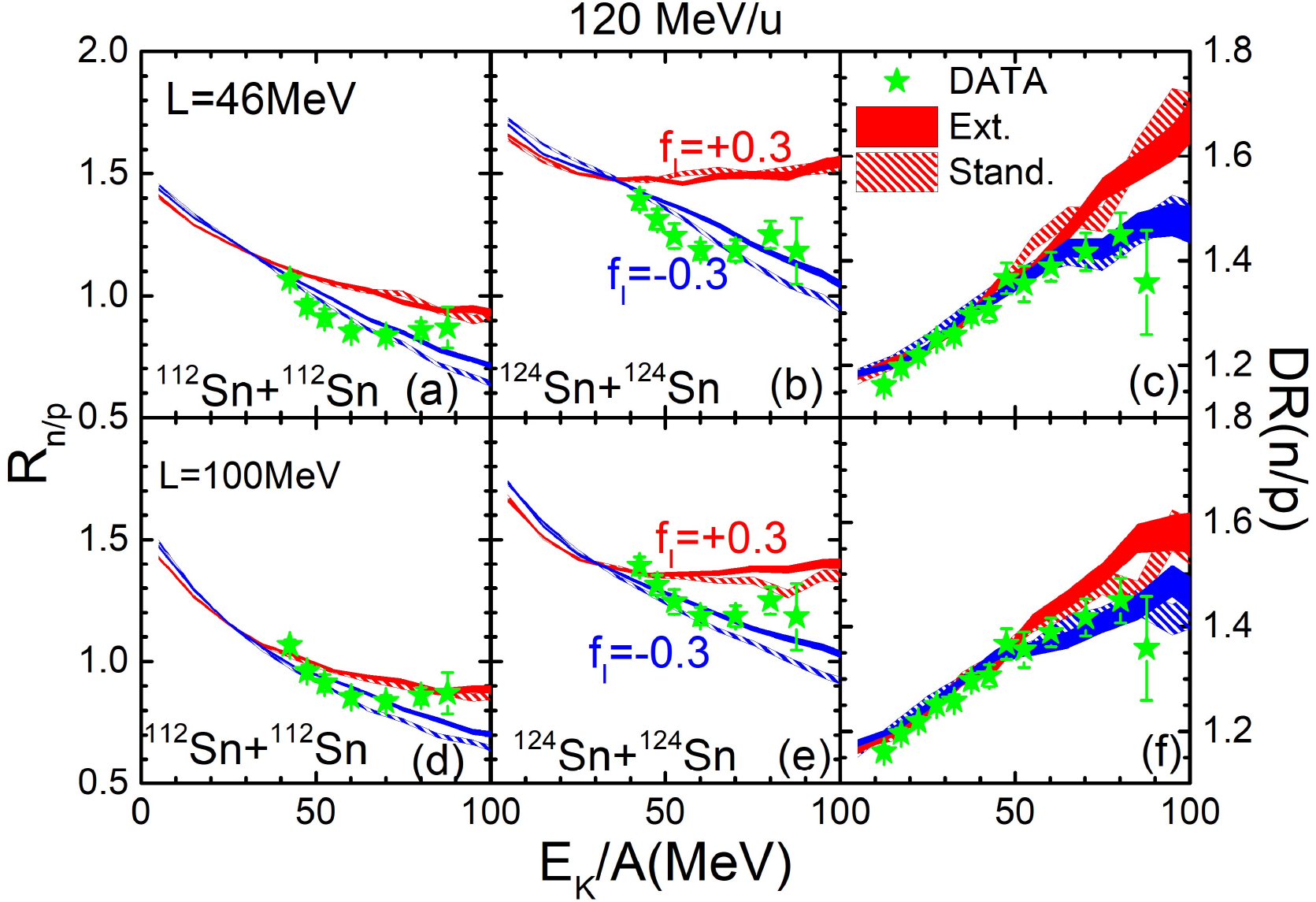}
\setlength{\abovecaptionskip}{0pt}
\vspace{0.2cm}
\caption{(Color online) The $R_{n/p}$ and $DR(n/p)$ as a function of $E_k/A$ at the beam energy $ E_\text{beam} = 120~\text{MeV/u} $ in transverse direction $70^{0}\leq\theta_{cm}\leq110^{0}$.}
\setlength{\belowcaptionskip}{0pt}
\label{NPratio-E120}
\end{figure}
\setlength{\abovedisplayskip}{3pt}
%$^{54}$Ni+$^{58}$Ni and $^{70}$Ni+$^{64}$Ni at the beam energy of 200 MeV/u. The $Y_{CI}(n)$ and $Y_{CI}(p)$ are obtained by combining the free nucleons with those bound in light isotopes with $1\leq A \leq5$\cite{coupland,Morfouace2019}. % we simulated the central collisions of $^{54}$Ni+$^{58}$Ni and $^{70}$Ni+$^{64}$Ni with incident energies of 200MeV per nucleon in the ImQMD model, where the collision parameter b=2, the total number of events is 50,000, and the dynamical evolution time is up to 200fm/c.

In Fig.~\ref{NPratio-E120}, we present the $R_{n/p}$ and $DR(n/p)$ as a function of $E_k/A$, i.e., the kinetic energy per nucleon of emitted particles in the center of mass frame at the beam energy $E_\text{beam} = 120~\text{MeV/u}$. The left and middle panels are $R_{n/p}$ for $^{112}$Sn+$^{112}$Sn and $^{124}$Sn+$^{124}$Sn, %for $^{54}$Ni+$^{58}$Ni and $^{70}$Ni+$^{64}$Ni, 
respectively. The right panels are for the $DR(n/p)$. %Panels (a) to (c) are $L$=46 MeV, and panels (d) to (f) are $L$= 100MeV. 
Our calculations show that the values of $R_{n/p}$ and $DR(n/p)$ obtained with $f_I=0.3$ are greater than that with $f_I=-0.3$ at high kinetic energy region for both MDIs (shaded regions) and both $L$ (the results for $L=46$ MeV are in the upper panels and the results for $L=100$ MeV are in the bottom panels). It can be understood from the symmetry potential presented in Fig.~\ref{fig:VLane}. %The solid shaded area is the result of the extended Skyrme-type MDI, while the dashed shaded area is the result of the standard Skyrme-type MDI. %The red shaded area represents $f_I=+0.3$ ( $m_n^*<m_p^*$), the blue shaded area represents $f_I=-0.3$ ($m_n^*>m_p^*$). 

Another, to understand the difference of the $R_{n/p}$ (or $DR(n/p)$) obtained with two MDIs, we also present the results obtained with standard Skyrme MDI in the dashed shaded regions. The values of $R_{n/p}$ obtained with the extended Skyrme-type MDI are almost the same as those obtained with the standard Skyrme MDI in the low kinetic energy region where the symmetry potential of the two MDI forms is basically the same. At the high kinetic energy region, the $R_{n/p}$ obtained with the extended Skyrme MDI is different than that with standard Skyrme MDI, but the difference depends on the $f_I$. In the case of $f_I=-0.3$, the $R_{n/p}$ obtained with extended Skyrme MDI are obviously larger than that with standard Skyrme MDI. But for $f_I=0.3$, the $R_{n/p}$ obtained with extended Skyrme MDI are close to that with standard Skyrme MDI. It seems contradictory with the symmetry potential presented in Fig.\ref{fig:VLane}, but can be understood from the reaction dynamics. 

\begin{figure}[htbp]
\flushleft
\includegraphics[width=\linewidth]{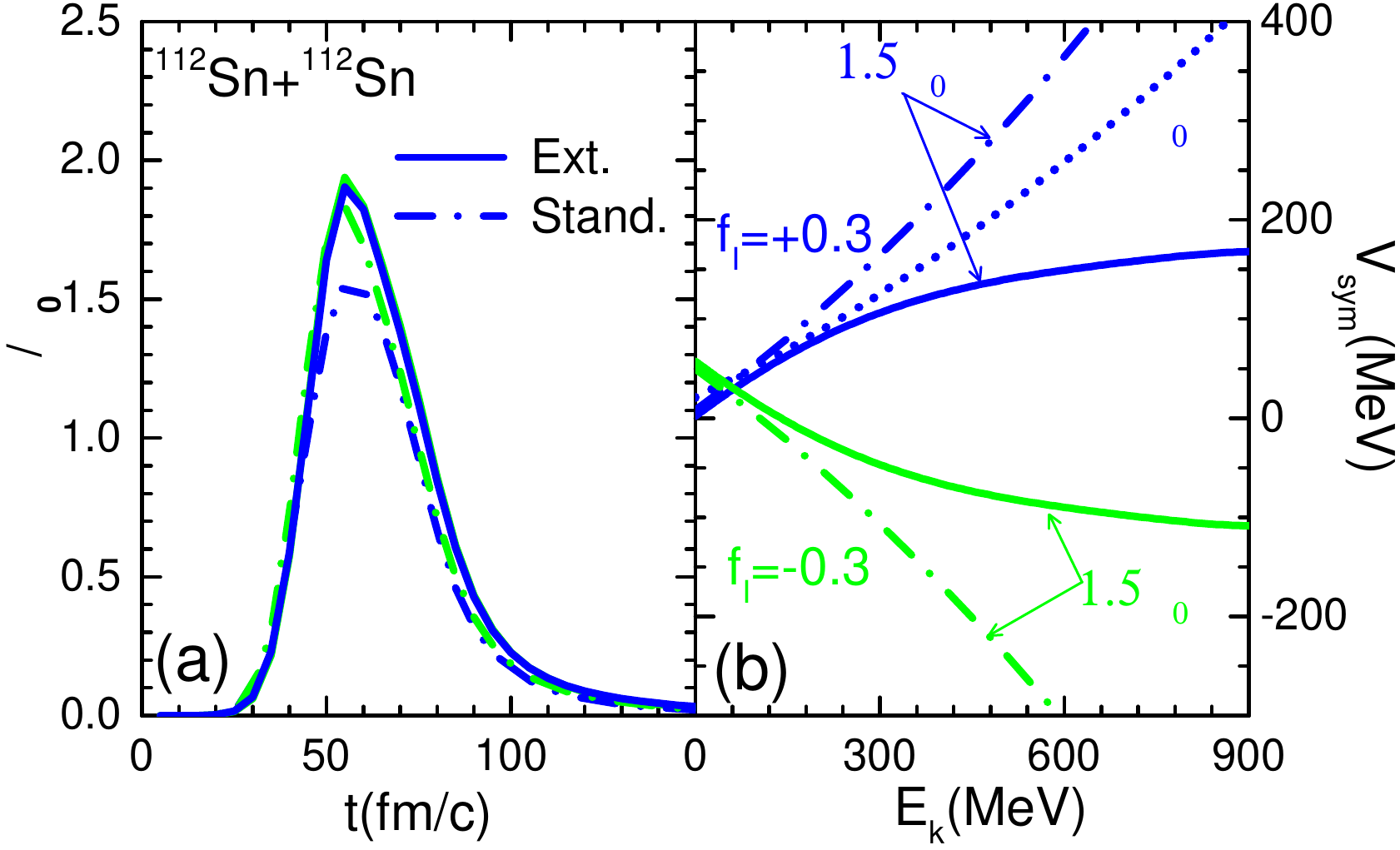}
\setlength{\abovecaptionskip}{0pt}
\vspace{0.2cm}
\caption{(Color online) Panel (a) is time evolution of average density in center of system; panel (b) is the symmetry potential for $f_I=0.3$ and $f_I=-0.3$ at different densities.}
\setlength{\belowcaptionskip}{0pt}
\label{fig:Vsym-rho}
\end{figure}

In the simulations of HICs, the maximum compressed density depends on the form of MDI as shown in Fig.\ref{fig:Vsym-rho} (a). For $f_I=-0.3$, our calculations show that the maximum compressed density reaches about 1.94$\rho_0$ for extended Skyrme MDI and reaches about 1.84$\rho_0$ for standard Skyrme MDI. The difference of the maximum compressed density obtained with two MDIs is less than 6\%. Thus, the difference of $R_{n/p}$ obtained with two MDIs will be similar with the difference of symmetry potential at a certain density. However, for $f_I=0.3$, the difference of the maximum compressed density obtained with two MDIs reaches about 18\%, where the maximum compressed density reaches about 1.9$\rho_0$ for extended Skyrme MDI and reaches about 1.56$\rho_0$ for standard Skyrme MDI. Consequently, the difference of $R_{n/p}$ obtained with two MDIs will not be similar with the difference of symmetry potential at a certain density. It will be like a difference of the symmetry potential at two densities for two MDIs. For example, as shown in Fig.\ref{fig:Vsym-rho} (b), the difference of the symmetry potential obtained with the extended Skyrme MDI at 1.5$\rho_0$ and the symmetry potential obtained with the standard Skyrme MDI at 1.0$\rho_0$ is small, and thus the difference of $R_{n/p}$ between two MDIs becomes smaller for $f_I=0.3$.

\begin{figure*}[htbp]
\flushleft
\includegraphics[width=\linewidth]{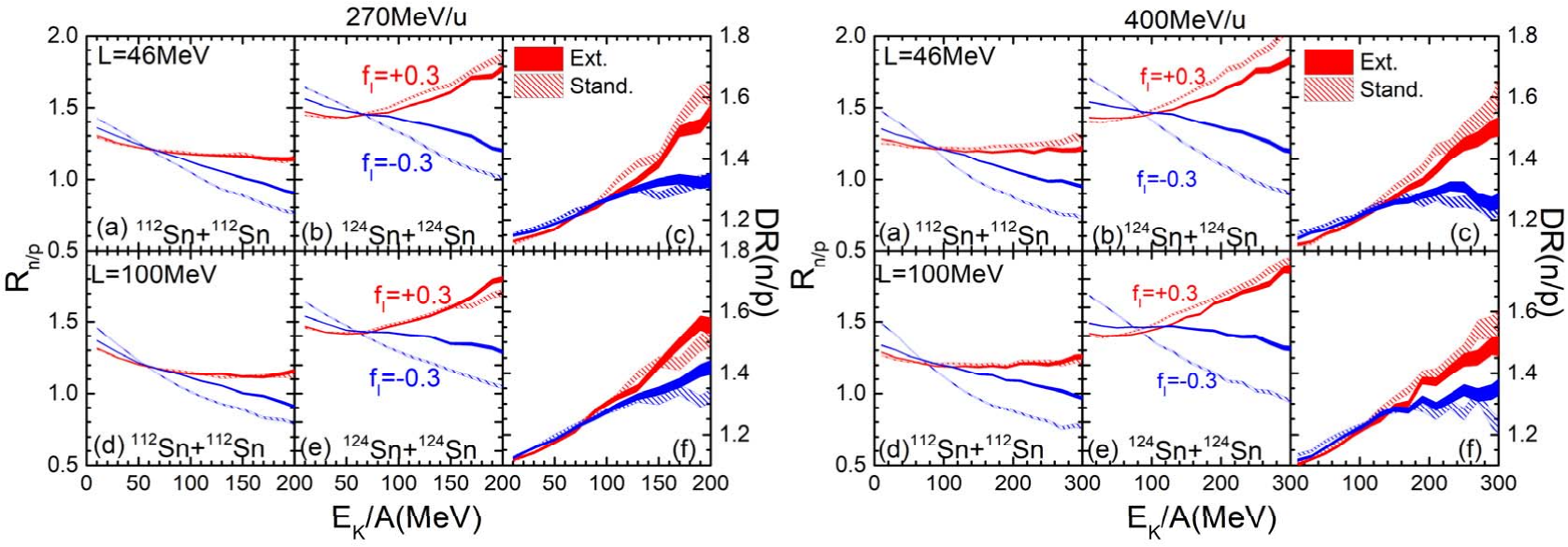}
\setlength{\abovecaptionskip}{0pt}
\vspace{0.2cm}
\caption{(Color online) The $R_{n/p}$ and $DR(n/p)$ as a function of $E_k/A$ at the beam energy $ E_\text{beam} = 270~\text{MeV/u},400~\text{MeV/u} $ in transverse direction $70^{0}\leq\theta_{cm}\leq110^{0}$.}
\setlength{\belowcaptionskip}{0pt}
\label{NPratio-E270400}
\end{figure*}

The differences of the single-particle potential between the extended Skyrme MDI and standard MDI become obvious at higher beam energies. Therefore, the $R_{n/p}$ and $DR(n/p)$ at the beam energies of 270 and 400 MeV/u, which is in the capability of ImQMD model, are also analyzed. The calculated results are presented in Figure~\ref{NPratio-E270400}. Similar to the results at 120 MeV/u, the $R_{n/p}$ and $DR(n/p)$ obtained with $f_I$=0.3 are greater than that with $f_I=-0.3$ in the high kinetic energy region.  %, but the difference of the $R_{n/p}$ and $DR(n/p)$ obtained with two $f_I$ increase a little at the beam energy region we studied.}
To quantitatively describe the isospin effective mass splitting effects on $R_{n/p}$, we construct the difference of $R_{n/p}$ between two kinds of $f_I$, i.e.,
\begin{equation}
    \Delta R_{n/p}=R_{n/p}(f_I=+0.3)-R_{n/p}(f_I=-0.3).
\end{equation}
Our calculations reveal that $\Delta R_{n/p}$ obtained with the extended Skyrme MDI is smaller than that with standard Skyrme MDI, and this difference increases with the beam energy.
%\textcolor{blue}{And we define the quantity A to represent the difference in $\Delta R_{n/p}$ between two different forms of MDI,}
% \begin{equation}
%     A=\frac{\Delta R_{n/p}(Stand.)-\Delta R_{n/p}(Ext.)}{\Delta R_{n/p}(Stand.)},
% \end{equation}
For example, for $^{124}$Sn+$^{124}$Sn, the difference of $\Delta R_{n/p}$ between the extended Skyrme MDI and standard Skyrme MDI increases
from 15\% to 45\% at $E_k/E_\text{beam}=0.75$ as the beam energy increases from $120~\text{MeV/u}$ to $400~\text{MeV/u}$.

\setlength{\abovedisplayskip}{3pt}

%Compared with the standard Skyrme-type MDI, the signle ratio of the CI neutron and proton $[\text{$Y(n)$}/\text{$Y(p)$}]_{CI}$ obtained by the extended Skyrme-type MDI is not significantly different in the low kinetic energy region, while the ratio of the CI neutron and proton obtained by the extended Skyrme-type MDI is larger in the high kinetic energy region when the effective mass splitting $f_{I}=-0.3$. 

%This is because the symmetry potential obtained by the standard Skyrme-type MDI and the extended Skyrme-type MDI is not significantly different in the low kinetic energy region and the expanded Skyrme-type MDI in the high kinetic energy region with $f_{I}=-0.3$ provides a stronger symmetric potential. But for $f_{I}=+0.3$ in the high kinetic energy region, there is little difference between the standard MDI and the extended MDI, the reason is that....

%Details of the deriviation of the terms of $\tilde{g}_{md}^{Ik}$ and $\tilde{C}_{sym}^{Ik}$ are referred in Appendix \ref{EOS-esky}.

\subsection{Discussions on the constraints of the effective mass splitting}

Further, we also compare the calculation to the corrected data of $R_{n/p}$ and $DR(n/p)$ (green symbols) that were published in Ref.~\cite{Morfouace2019}. In general, the calculated $R_{n/p}$ and $DR_{n/p}$ obtained with $f_I=-0.3 (\Delta m_{np}^*\approx 0.311 \delta)$ are close to the data points for $L=46$ MeV. In the case of $L=100$ MeV, the data falls into the middle between the results obtained with $f_I=-0.3 (\Delta m_{np}^*\approx 0.311 \delta)$ and $f_I=0.3 (\Delta m_{np}^*\approx -0.311 \delta)$. It implies that the $\Delta m_{np}^*$ is negatively correlated to the $L$ and which is consistent with the results obtained with Hugenholtz-Van Hove (HVH) theorem ~\cite{ChangXu2010PRC}. This conclusion is consistent with the previous results in Ref.~\cite{Morfouace2019}. However, one should note that the $DR(n/p)$ suppresses the sensitivity to effective mass splitting, and the curves of $R_{n/p}$ as a function of $E_k$ are different than the data for both reaction systems if we carefully check its shape. Therefore, it is important to further quantitatively analyze the shapes of $R_{n/p}$ as a function of $E_k$.
%A firm conclusion on the effective mass splitting needs a careful comparison of $R_{n/p}$ and $DR(n/p)$ to data.

To probe the strength of the effective mass splitting which only depends on the momentum-dependent part of the symmetry potential [as mentioned in Eq.(\ref{eq:dmn-vasy})], one has to single out the contributions from the momentum-dependent part of the symmetry potential. The slope of $\ln{R_{n/p}}$ as a function of $E_k/A$, i.e., 
\begin{equation}
S_{n/p}=\frac{\partial \ln{R_{n/p}}}{\partial E_k/A},    
\end{equation}
can be used as mentioned in Ref.~\cite{FYWang2023NST}. %In the case of standard Skyrme interaction, %The reason is that $S_{n/p}$ has been verified to be directly sensitive to the strength of effective mass splitting~\cite{FYWang2023} in the case of the standard 
According to the statistical and dynamical model~\cite{Tsang2001prl, Tsang2001prc, Ono2003, Das1981, George1987, Botvina2002}, the neutron to proton yield ratios $R_{n/p}$ can be written as, %The reason can be understood in the statistics model \cite{Tsang2001prl, Tsang2001prc, Ono2003, Das1981, George1987, 
\begin{equation} \label{ynyp-vlane}
\begin{aligned}
   R_{n/p}= \frac{\text{Y(n)}}{\text{Y(p)}}&\propto \exp \left( \frac{ \mu_n - \mu_p}{ T }\right)=\exp \left( \frac{ 2V_{sym}\delta}{ T }\right).
\end{aligned}
\end{equation}
%\begin{equation} \label{ynyp-vlane}
%\begin{aligned}
%    \frac{\text{Y(n)}}{\text{Y(p)}} &\propto \exp \left( \frac{ \mu_n - \mu_p}{ T }\right) \\
%   &=\exp \Bigg[V_{sym}^{loc}/T+2\tilde{D}_0\sum_{I=1}^N b_{I}\sum_{l=0,l\in even}^{2I}\mathcal{\Tilde{A}}_{Il}\\
%   &\quad \left(\frac{\rho}{2}\right)^{(2I-l+3)/3} (2\times\frac{2I-l+3}{3}\delta+\dots) p^l/T\Bigg]\\
%   &=\exp \Bigg[\frac{V_{sym}^{loc}}{T}+f_{I}\times \frac{2E_{k}}{lT}(2\times\frac{2I-l+3}{3}\delta+\dots)\Bigg].
%\end{aligned}
%\end{equation}
%here,$V_{sym}^{loc}$=$4A_\text{sym} \frac{\rho}{\rho_0 }\delta$ + $4B_\text{sym}\left( \frac{\rho}{\rho_0 }\right)^{\gamma}\delta$.
$T$ is the temperature of the emitting source, $\mu_n$ and $\mu_p$ are the chemical potentials of neutrons and protons, respectively. If we expand the symmetry potential with respect to $p^2$, i.e., $V_{sym}=V_{sym}^0+\frac{\partial V_{sym}}{\partial p^2}p^2+\cdots$, then $R_{n/p}$ can be written as,
\begin{equation} \label{eq:Rnp-Vasy-ext}
\begin{aligned}
   R_{n/p}&\propto \exp \Bigg[\frac{2(V_{sym}^0+\frac{\partial V_{sym}}{\partial p^2}p^2+\cdots )\delta}{T}\Bigg]\\
   &\approx\exp \Bigg[\frac{2V_{sym}^0\delta}{T}\Bigg] \exp \Bigg[-\frac{(\frac{m}{m^*})^2\Delta m_{np}^*}{T}E'_k\Bigg].
\end{aligned}
\end{equation}
$E'_k=p_{rel}^2/2m$ is the relative kinetic energy between colliding nucleon pairs during the collisions, which should positively correlate to the kinetic energy of emitted nucleons. If we simply assume $E'_k=\lambda E_k/A$, then we have
\begin{equation}
S_{n/p}=-\frac{\lambda}{T}(\frac{m}{m^*})^2\Delta m_{np}^*,
\end{equation}
which is directly related to the $\Delta m_{np}^*$. If one assumes $\lambda=1$ and $T=5$ MeV, the estimated values of $S_{n/p}$ are in the range from $\pm 0.02$ for the parameters we used, i.e., $m_s^*/m=0.77$ and $f_I=\pm 0.3$. In the HICs, the $\lambda$ is related to the friction of the system and the value should be smaller than one. Thus, one can expect the $|S_{n/p}|<0.02$ if $T=5$ MeV. 

\begin{figure}[htbp]
\flushleft
\includegraphics[width=\linewidth]{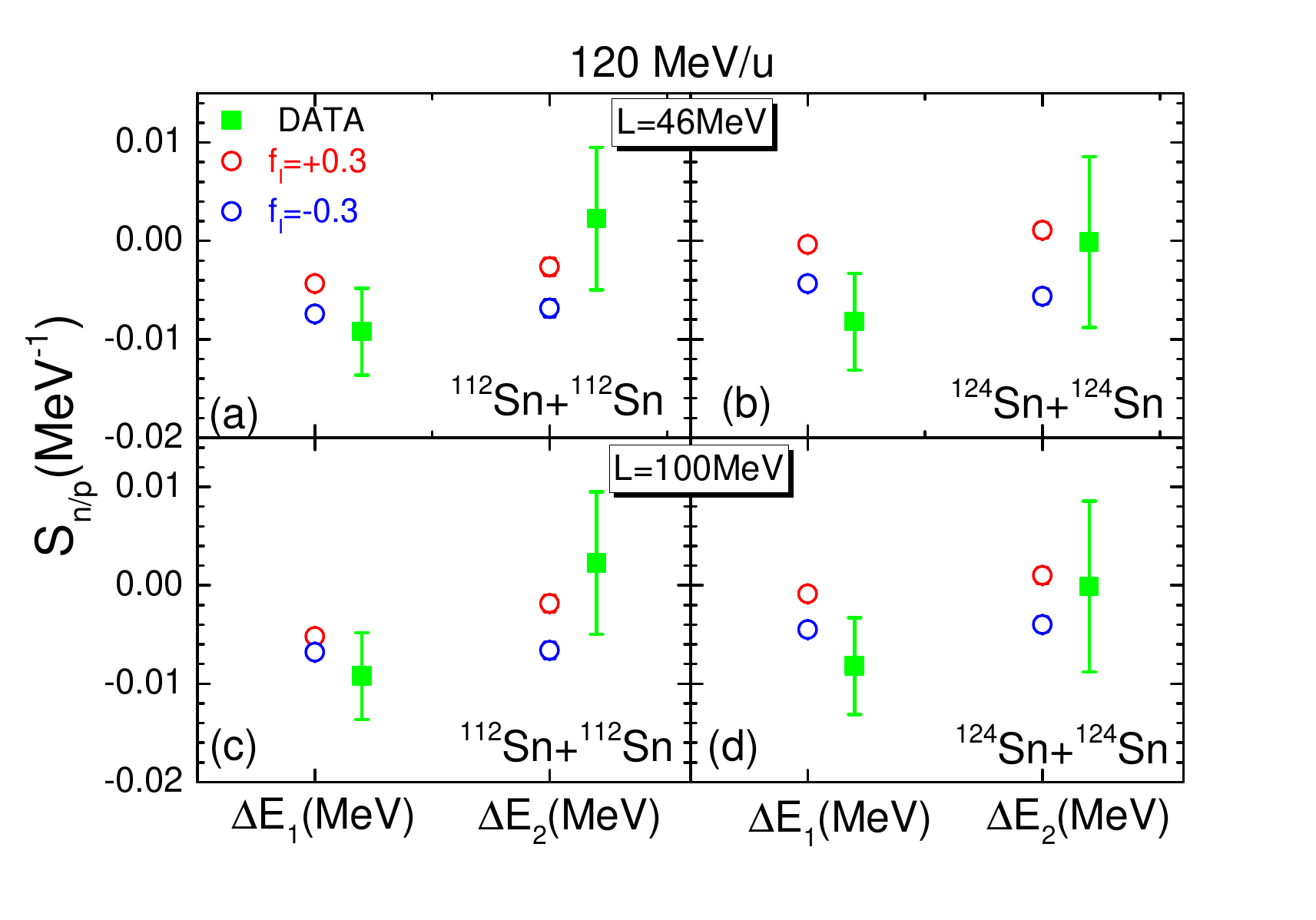}
\setlength{\abovecaptionskip}{0pt}
\vspace{0.2cm}
\caption{(Color online). The $S_{n/p}$ at two kinetic regions and at the beam energy $ E_\text{beam} = 120~\text{MeV/u} $ in the transverse direction $70^{0}\leq\theta_{cm}\leq110^{0}$. Data points are extracted from the published $R_{n/p}$ in Ref.~\cite{Morfouace2019}.}
\setlength{\belowcaptionskip}{0pt}
\label{fig:Snp}
\end{figure}
\setlength{\abovedisplayskip}{3pt}

In Fig.\ref{fig:Snp}, we present the $S_{n/p}$ obtained in the simulation and data. Two kinetic regions, i.e., the kinetic energy per nucleon of the emitted particles in the range of 35MeV $\leq \Delta E_1\leq$ 55MeV and 55MeV $\leq \Delta E_2 \leq$ 95MeV, are used to calculate $S_{n/p}$. The systems of $^{112}$Sn+$^{112}$Sn and $^{124}$Sn+$^{124}$Sn at the beam energy of 120 MeV/u are presented in the left and right panels, respectively. The red circles represent the results obtained with $f_I=0.3$, the blue circles represent the results with $f_I=-0.3$. The green squares are the data points of $S_{n/p}$ which are extracted from the published experimental data~\cite{Morfouace2019}.
One can find that the calculations tend to favor different effective mass splitting at different kinetic energy regions. In quality, the calculated values of $S_{n/p}$ tend to favor $f_I<0$ ($m_n^*>m_p^*$) at the kinetic energy less than 55 MeV. At the kinetic energy greater than 55 MeV, the calculated values of $S_{n/p}$ tend to favor $f_I>0$ ($m_n^*<m_p^*$). It implies the $m_n^*>m_p^*$ at low kinetic energy and $m_n^*<m_p^*$ at high kinetic energy, which is consistent with the theoretical calculations from the relativistic Hartree-Fock calculations\cite{WHLong}. In quantity, the calculations can not exactly describe the shape of $R_{n/p}$ by the parameter sets we used. The possible reasons may be a reasonable parameter set is not used or the cluster formation for light particles is not well described. To rule out the first reason, a systematic analysis of $R_{n/p}$ on the multi-dimensional parameter space is certainly needed before going to understand the cluster formation mechanism and draw a firm conclusion on the effective mass splitting.

In addition, the uncertainties of the transport models should also be investigated in future studies. Generally, the uncertainty of the transport models arises from three aspects. One is from the numerical techniques, and another is from the uncertainty of the physical quantities. The third one is from the systematic uncertainties due to the missed physics. The first point has stimulated the transport model evaluation project (TMEP) to improve the transport model. There are some important progresses have been made in the mean-field part ~\cite{MariaPRC2021}, nucleon-nucleon collision part~\cite{YXZhangPRC2018,Ono2019PRC}, and the recent progress is suggested to read the Ref.~\cite{Hermann2022PPNP}. The second point has led to the application of the Bayesian analysis on the heavy ion collision observables in multi-dimensional parameter space as in ~\cite{Morfouace2019,YXZhang20}. One should notice that the new form of the extended MDI may influence the in-medium nucleon-nucleon cross sections from the point view of both mean field and the in-medium nucleon-nucleon cross sections in the transport equation have the same origin. When the in-medium nucleon-nucleon cross sections are modified in the transport model, the values of $R_{n/p}$ may be further modified except for the impacts from the mean field potential. Nonetheless, the results in Ref.~\cite{YXZhangPRC2012} suggest that the medium correction on the nucleon-nucleon cross sections weakly influences the $R_{n/p}$ since it is mainly determined by the strength of symmetry energy. To quantitatively estimate the uncertainty of this medium correction of the nucleon-nucleon cross sections on $R_{n/p}$, a Bayesian analysis of the heavy ion collisions observables in multi-dimensional parameter space, which includes the degree of the medium corrections on the nucleon-nucleon cross sections, is needed. The third one comes from the philosophy of solving the many-body transport equations and missed physics, which will be an important point to develop an advanced model for describing the complex collisions as wished in the long-range plan of nuclear science~\cite{2023LRP}.

\section{SUMMARY}
\label{sec:summary}
The main goal of this work is to obtain an extended Skyrme momentum-dependent interaction, and its related mean-field potential and Hamiltonian. The forms we provided have the following properties. %2) the parameters in the extended Skyrme momentum-dependent interaction in our work are determined by fitting the Hama's optical potential data as well as the nuclear matter parameters. Two parameters were added in front of extended Skyrme MDI, 
%and is flexible to adjust the strength of the momentum-dependent symmetry potential to investigate the effective mass splitting in transport models. 
First, the extended Skryme MDI extends the utility of the effective Skyrme interaction to a wider energy region in the transport models. For instance, this form can be used to calculate the vector potential and scalar potential for studying the effects of the effective mass splitting and the isospin-dependent threshold of the pion production for HICs below 1 GeV/u as in Ref.~\cite{zhangzhenPRC2018}. Second, the effect of the width of the wave packet can be explicitly involved in the MDI part of the QMD-type Hamiltonian. Third, this form can be easily used in both the BUU and QMD-type models for simulating the heavy ion collisions at intermediate to high energies. In another, the relationship between the nuclear matter parameters and the interaction parameters used in the transport models are provided, which makes the transport model investigate the nuclear matter parameter in multi-dimensional parameter space easily.
%\textcolor{red}{The advantages of the extended MDI are in two factors. One is that this form can treat the effects of the width of the wave packet explicitly in the QMD-type Hamiltonian. Another point is that this form can be used in the wider energy range than the standard Skyrme interaction, for example, this form can be easily used to calculate the vector potential and scalar potential for studying the effective mass splitting effects and the threshold effects of the pion production as in Ref.\cite{zhangzhenPRC2018}.}

%In summary, we present an extended Skyrme momentum-dependent interaction which can fit the Hama optical potential and can be analytically incorporated in the QMD type models. We further extend this form to an isospin dependent momentum-dependent interaction. Based on the extended Skyrme momentum-dependent interaction, we can get the isospin dependent single-particle potential and the Hamiltonian with Gaussian wave packets analytically, which will be useful in the BUU and QMD type models.

As an example of the application of the extended Skyrme MDI, we incorporate the extended Skyrme MDI into the ImQMD model. The isospin sensitive observables, such as the single and double ratios of the coalescence neutron to proton yields ($R_{n/p}$ and $DR(n/p)$), for $^{112,124}$Sn+$^{112,124}$Sn at the beam energy of 120 MeV/u, 270 MeV/u, and 400 MeV/u are analyzed to understand the influence of effective mass splitting. Our calculations show that the difference of the $R_{n/p}$ ratios obtained with two different strengths of the effective mass splitting, i.e., $f_I=+0.3$ and $f_I=-0.3$, becomes weaker for the extended Skyrme MDI than that for the standard Skyrme MDI.% \textcolor{red}{At high beam energy, the difference of the $R_{n/p}$ ratios obtained with two different $f_I$ becomes stronger.}

Finally, we also propose a probe for constraining the strength of the effective mass splitting, i.e., the slope of $\ln{R_{n/p}}$ as a function of $E_k/A$, $S_{n/p}=\frac{\partial \ln{R_{n/p}}}{\partial E_k/A}$. Our calculations show that $S_{n/p}$ is directly related to the effective mass splitting $\Delta m_{np}^*$. By comparing $S_{n/p}$ with the experimental data, in quality, one can find that the calculations favor different effective mass splitting at different kinetic energy regions. The calculated values of $S_{n/p}$ tend to favor $m_n^*>m_p^*$ at the kinetic energy less than 55 MeV and tend to favor $m_n^*<m_p^*$ at the kinetic energy greater than 55 MeV. In quantity, the calculations can not accurately describe the shape of $R_{n/p}$ by the parameter sets we used. Thus, further analysis on the $R_{n/p}$ in multi-dimensional parameter space is certainly needed to understand the discrepancy and to reliably constrain the effective mass splitting. %for different reaction systems and different beam energies 

%In our work, we apply the extended Skyrme momentum-dependent interaction to the ImQMD model to see the effects of the extended Skyrme MDI on the HICs observables under different the solpes $L$ and effective mass splitting $f_I$. The single ratio of the coalescence invariant (CI) neutron and proton $R_{n/p}=Y_{CI}(n)/Y_{CI}(p)$ and the double ratio of the coalescence invariant (CI) neutron and proton $DR(n/p)=R_{n/p}(A)/R_{n/p}(B)$ are analyzed for the central collisions of $B=^{112}$Sn+$^{112}$Sn and $A=^{124}$Sn+$^{124}$Sn at the beam energy of 120 MeV/u. Our calculations show than the slope of $\ln{R_{n/p}}$ as a function of $E_k/A$, i.e., $S_{n/p}=\frac{\partial \ln{R_{n/p}}}{\partial E_k/A}$ is directly related to the $\Delta m_{np}^*$. By comparing $S_{n/p}$ with the experimental data, one can find that the calculations favor different effective mass splitting at different kinetic energy regions. The calculated values of $S_{n/p}$ favor $f_I<0$ ($m_n^*>m_p^*$) at low kinetic energy, at the kinetic energy less than 50 MeV and $f_I>0$ ($m_n^*<m_p^*$) at the kinetic energy greater than 50 MeV. In quantity, the calculations can not accurately describe the shape of $R_{n/p}$ by the parameter sets we used, therefore by analyzing the slope of $\ln{R_{n/p}}$ in a multi-dimensional parameter space is certainly needed to further constrain effective mass splitting.

\section*{Acknowledgments}

This work was partly inspired by the transport model evaluation project, and it was supported by the National Natural
Science Foundation of China under Grants No. 12275359, No. 12375129, No. 11875323 and No. 11961141003, by the National Key R\&D Program of China under Grant No. 2023 YFA1606402, by the Continuous Basic Scientific Research Project, by funding of the China Institute of Atomic Energy under Grant No. YZ222407001301, No. YZ232604001601, and by the Leading Innovation Project of the CNNC under Grants No. LC192209000701 and No. LC202309000201.

\appendix
\section{Energy density of the extended Skyrme-type momentum-dependent interaction}\label{appendix:EOS-esky}

Extended Skyrme-type MDI energy density is
\begin{eqnarray} \label{EextMDI}
\begin{aligned}
    u_{md} & = \tilde{C}_0\int \text{d}^3p\text{d}^3p' 
    %(f_n(\mathbf{r},\mathbf{p})+f_p(\mathbf{r},\mathbf{p}))(f_n(\mathbf{r},\mathbf{p'})\\
    %& \quad +f_p(\mathbf{r},\mathbf{p'}))g(\mathbf{p}-\mathbf{p'}) \\
    f(\mathbf{r},\mathbf{p})f(\mathbf{r},\mathbf{p'})g(\mathbf{p}-\mathbf{p'}) \\
    & \quad + \tilde{D}_0\int \text{d}^3 p \text{d}^3p' f_n(\mathbf{r},\mathbf{p}) f_n(\mathbf{r},\mathbf{p'})g(\mathbf{p}-\mathbf{p'}) \\
    & \quad + \tilde{D}_0\int \text{d}^3p \text{d}^3p' f_p(\mathbf{r},\mathbf{p}) f_p(\mathbf{r},\mathbf{p'})g(\mathbf{p}-\mathbf{p'}). \\
    & = \tilde{C}_0\int \text{d}^3p\text{d}^3p'\\
    &\quad \Bigg(f_n(\mathbf{r},\mathbf{p})f_n(\mathbf{r},\mathbf{p'})+f_n(\mathbf{r},\mathbf{p})f_p(\mathbf{r},\mathbf{p'})\\
    & \quad +f_p(\mathbf{r},\mathbf{p})f_n(\mathbf{r},\mathbf{p'})+f_p(\mathbf{r},\mathbf{p})f_p(\mathbf{r},\mathbf{p'})\Bigg)g(\mathbf{p}-\mathbf{p'}) \\
    & \quad + \tilde{D}_0\int \text{d}^3 p \text{d}^3p' f_n(\mathbf{r},\mathbf{p}) f_n(\mathbf{r},\mathbf{p'})g(\mathbf{p}-\mathbf{p'}) \\
    & \quad + \tilde{D}_0\int \text{d}^3p \text{d}^3p' f_p(\mathbf{r},\mathbf{p}) f_p(\mathbf{r},\mathbf{p'})g(\mathbf{p}-\mathbf{p'}). \\
%    & = u_{md}^{sym}+ u_{md}^{asym}\delta^2.
\end{aligned}
\end{eqnarray}
and $g(\mathbf{p}-\mathbf{p'})$ is taken as,
\begin{eqnarray} \label{gp1p2}
\begin{aligned}
    g(\mathbf{p}-\mathbf{p'}) &= b_{0}+b_{1}(\mathbf{p}-\mathbf{p'})^{2}+b_{2}(\mathbf{p}-\mathbf{p'})^{4}\\
    &\quad +b_{3}(\mathbf{p}-\mathbf{p'})^{6}+b_{4}(\mathbf{p}-\mathbf{p'})^{8}\\
    =&\sum_{I=0}^{N=4} b_{I}\Bigg[ \sum_{l=0,\in even}^{2I}\tbinom{2I}{l} p^l (-p')^{2I-l}\\
    & \quad \quad -\sum_{l=1,\in odd}^{2I}\tbinom{2I}{l} p^{l-1} (-p')^{2I-l-1}\mathbf{p}\cdot \mathbf{p}' \Bigg].\\
%    &+b_2\sum_{k=0}^2 C_2^k\mathbf{p}^k\cdot(-\mathbf{p'})^{n-k}+b_3\sum_{k=0}^3 C_2^k\mathbf{p}^k\cdot(-\mathbf{p'})^{n-k}\\
%    &+b_4\sum_{k=0}^4 C_2^k\mathbf{p}^k\cdot(-\mathbf{p'})^{n-k}
%    =&b_{0}+b_{1}(\mathbf{p}^2-2\mathbf{p}\cdot\mathbf{p'}+\mathbf{p'}^2)\\
%    &+b_{2}(\mathbf{p}^4-4\mathbf{p}^3\cdot\mathbf{p'}+6\mathbf{p}^2\mathbf{p'}^2-4\mathbf{p}\cdot\mathbf{p'}^3+\mathbf{p'}^4)\\
%    &+b_{3}(\mathbf{p}^6-6\mathbf{p}^5\cdot\mathbf{p'}+15\mathbf{p}^4\mathbf{p'}^2-20\mathbf{p}^3\cdot\mathbf{p'}^3\\
%    &+15\mathbf{p}^2\mathbf{p'}^4-6\mathbf{p}\cdot\mathbf{p'}^5+\mathbf{p'}^6)\\
%    &+b_{4}(\mathbf{p}^8-8\mathbf{p}^7\mathbf{p'}+28\mathbf{p}^6\mathbf{p'}^2-56\mathbf{p}^5\mathbf{p'}^3\\
%    &+70\mathbf{p}^4\mathbf{p'}^4-56\mathbf{p}^3\mathbf{p'}^5+28\mathbf{p}^2\mathbf{p'}^6-8\mathbf{p}\mathbf{p'}^7+\mathbf{p'}^8).
\end{aligned}
\end{eqnarray}
For cold uniform nuclear matter, $f_q=\frac{2}{(2\pi\hbar)^3}\Theta(p_{F_q}-p)$, q=n/p, and analytical expression of $u_{md}$ can be obtained.

For the $\tilde{C}_0$ term,
%(1) For isospin-like term, i.e., $q$=$q'$=n or p, $I_1=C_0\int \text{d}^3p\text{d}^3p' f_q(\mathbf{r},\mathbf{p})f_{q'}(\mathbf{r},\mathbf{p'})g(\mathbf{p}-\mathbf{p'})$,
%with $q=q'$=n or p, %    \\ $\ C_0\int \text{d}^3p\text{d}^3p' f_p(\mathbf{r},\mathbf{p})f_p(\mathbf{r},\mathbf{p'})g(\mathbf{p}-\mathbf{p'})$
\begin{equation}
    \begin{aligned}
&\tilde{C}_0\int \text{d}^3p\text{d}^3p' \bigg[f_n(\mathbf{r},\mathbf{p})f_n(\mathbf{r},\mathbf{p'})+f_n(\mathbf{r},\mathbf{p})f_p(\mathbf{r},\mathbf{p'})\\
& \quad +f_p(\mathbf{r},\mathbf{p})f_n(\mathbf{r},\mathbf{p'})+f_p(\mathbf{r},\mathbf{p})f_p(\mathbf{r},\mathbf{p'})\Bigg]g(\mathbf{p}-\mathbf{p'}) \\
&=\tilde{C}_0\left(\frac{2}{(2\pi\hbar)^3}\right)^2 \Bigg(\int_0^{p_{F_n}} \text{d}\mathbf{p} \int_{0}^{p_{F_n}}\text{d}\mathbf{p'}\\
& \quad +\int_0^{p_{F_n}} \text{d}\mathbf{p} \int_{0}^{p_{F_p}}\text{d}\mathbf{p'}+\int_0^{p_{F_p}} \text{d}\mathbf{p} \int_{0}^{p_{F_n}}\text{d}\mathbf{p'}\\
& \quad +\int_0^{p_{F_p}} \text{d}\mathbf{p} \int_{0}^{p_{F_p}}\text{d}\mathbf{p'}\Bigg)\\
&\quad \times \sum_{I=0}^4 b_{I}\Bigg[ \sum_{l=0,\in even}^{2I}\tbinom{2I}{l} p^l (-p')^{2I-l}\\
    & \quad \quad \quad -\sum_{l=1,\in odd}^{2I}\tbinom{2I}{l} p^{l-1} (-p')^{2I-l-1}\mathbf{p}\cdot \mathbf{p}' \Bigg].\\
%&=\tilde{C}_0\left(\frac{2}{(2\pi\hbar)^3}\right)^2(4\pi)^2 \Bigg[\sum_{I=0}^N b_{I}\sum_{l=0,l\in even}^{2I}\tbinom{2I}{l}\\
%&\quad(\int_0^{p_{F_n}} p^{2+l}\text{d}p \int_{0}^{p_{F_n}}p'^{2+2I-l}\text{d}p'+\int_0^{p_{F_n}} p^{2+l}\text{d}p \\
%&\quad\int_{0}^{p_{F_p}}p'^{2+2I-l}\text{d}p'+\int_0^{p_{F_p}} p^{2+l}\text{d}p \int_{0}^{p_{F_n}}p'^{2+2I-l}\text{d}p'\\
%&\quad+\int_0^{p_{F_p}} p^{2+l}\text{d}p \int_{0}^{p_{F_p}}p'^{2+2I-l}\text{d}p') \Bigg]\\
%&=\tilde{C}_0\left(\frac{2}{(2\pi\hbar)^3}\right)^2(4\pi)^2\\
%&\quad \times\Bigg[\sum_{I=0}^N b_{I}\sum_{l=0,l\in even}^{2I}\tbinom{2I}{l}\frac{1}{(l+3)}\frac{1}{(2I-l+3)}\\
%&\quad \times \bigg(p_{F_n}^{l+3}p_{F_n}^{2I-l+3}+p_{F_n}^{l+3}p_{F_p}^{2I-l+3}\\
%&\quad \quad \quad \quad  +p_{F_p}^{l+3}p_{F_n}^{2I-l+3}+p_{F_p}^{l+3}p_{F_p}^{2I-l+3}\bigg)\Bigg]\\
%&\quad =\tilde{C}_0\Bigg[\sum_{I=0}^N b_{I}\sum_{l=0,l\in even}^{2I}\mathcal{G}_{Il} \sum_{q,q'}\rho_{q}^{(l+3)/3}\rho_{q'}^{(2I-l+3)/3}\Bigg].
    \end{aligned}
\end{equation}

Since the terms with $l$ equal an odd number are zero, the $\tilde{C}_0$ term will be,
\begin{equation}
    \begin{aligned}
%&C_0\left(\frac{4}{(2\pi\hbar)^3}\right)^2(4\pi)^2\int_0^{p_F} p^2\text{d}p \int_{0}^{p_F}p'^2\text{d}p' \\
%&\quad\quad\big(\sum_{I=0}^4 b_{I}\sum_{k=0}^{2I}\tbinom{2I}{k}\mathbf{p}^k\cdot(-\mathbf{p'})^{2I-k}\big)\\
&=\tilde{C}_0\left(\frac{2}{(2\pi\hbar)^3}\right)^2(4\pi)^2 \Bigg[\sum_{I=0}^N b_{I}\sum_{l=0,l\in even}^{2I}\tbinom{2I}{l}\\
&\quad(\int_0^{p_{F_n}} p^{2+l}\text{d}p \int_{0}^{p_{F_n}}p'^{2+2I-l}\text{d}p'+\int_0^{p_{F_n}} p^{2+l}\text{d}p \\
&\quad\int_{0}^{p_{F_p}}p'^{2+2I-l}\text{d}p'+\int_0^{p_{F_p}} p^{2+l}\text{d}p \int_{0}^{p_{F_n}}p'^{2+2I-l}\text{d}p'\\
&\quad+\int_0^{p_{F_p}} p^{2+l}\text{d}p \int_{0}^{p_{F_p}}p'^{2+2I-l}\text{d}p') \Bigg]\\
&=\tilde{C}_0\left(\frac{2}{(2\pi\hbar)^3}\right)^2(4\pi)^2\\
&\quad \times\Bigg[\sum_{I=0}^N b_{I}\sum_{l=0,l\in even}^{2I}\tbinom{2I}{l}\frac{1}{(l+3)}\frac{1}{(2I-l+3)}\\
&\quad \times \bigg(p_{F_n}^{l+3}p_{F_n}^{2I-l+3}+p_{F_n}^{l+3}p_{F_p}^{2I-l+3}\\
&\quad \quad \quad \quad  +p_{F_p}^{l+3}p_{F_n}^{2I-l+3}+p_{F_p}^{l+3}p_{F_p}^{2I-l+3}\bigg)\Bigg]\\
&\quad =\tilde{C}_0\Bigg[\sum_{I=0}^N b_{I}\sum_{l=0,l\in even}^{2I}\mathcal{G}_{Il} \sum_{q,q'}\rho_{q}^{(l+3)/3}\rho_{q'}^{(2I-l+3)/3}\Bigg]   
%&=C_0\left(\frac{4}{(2\pi\hbar)^3}\right)^2(4\pi)^2 \Bigg[\\
%&\quad \quad b_1(\frac{1}{15}+\frac{1}{15})p_F^{8}\\
%&\quad +b_2(\frac{1}{21}+\frac{6}{25}+\frac{1}{21})p_F^{10}\\
%&\quad +b_3(\frac{1}{27}+\frac{15}{35}+\frac{15}{35}+\frac{1}{27})p_F^{12}\\
%&\quad +b_4(\frac{1}{33}+\frac{28}{45}+\frac{70}{49}+\frac{28}{45}+\frac{1}{33})p_F^{14}\Bigg]\\
    \end{aligned}
\end{equation}

Here, 
\begin{equation*}
\begin{aligned}
\mathcal{G}_{Ik}&=\left(\frac{2}{(2\pi\hbar)^3}\right)^2(4\pi)^2\Bigg[\left(\hbar(3\pi^2)^{1/3}\right)^{(2I+6)}\\
&\quad \tbinom{2I}{l}\frac{1}{(l+3)}\frac{1}{(2I-l+3)}\Bigg],    
\end{aligned}
\end{equation*}
and $p_{F_q}=\hbar(3\pi^2\rho_q)^{1/3}$.

Similarly, the $\tilde{D}_0$ terms,
\begin{equation}
    \begin{aligned}
&\tilde{D}_0\int \text{d}^3p\text{d}^3p' f_q(\mathbf{r},\mathbf{p})f_q(\mathbf{r},\mathbf{p'})g(\mathbf{p}-\mathbf{p'}) \\
&=\tilde{D}_0\left(\frac{2}{(2\pi\hbar)^3}\right)^2(4\pi)^2\int_0^{p_{F_q}} p^2\text{d}p \int_{0}^{p_{F_q}}p'^2\text{d}p' \\
&\quad\quad\big(\sum_{I=0}^N b_{I}\sum_{l=0,l\in even}^{2I}\tbinom{2I}{l}p^l(-p')^{2I-l}\big)\\
&=\tilde{D}_0\left(\frac{2}{(2\pi\hbar)^3}\right)^2(4\pi)^2 \Bigg[\sum_{I=0}^N b_{I}\sum_{l=0,l\in even}^{2I}\tbinom{2I}{l}\\
&\quad \quad \frac{1}{(l+3)}\frac{1}{(2I-l+3)}p_{F_q}^{2I+6}\Bigg]\\
&\quad =\tilde{D}_0\Bigg[\sum_{I=0}^N b_{I}\sum_{l=0,l\in even}^{2I} \mathcal{G}_{Il} \times \rho_{q}^{(2I+6)/3}\Bigg].
    \end{aligned}
\end{equation}

The $u_{md}$ will be,
\begin{equation}
    \begin{aligned}
u_{md}&=\tilde{C}_0\Bigg[\sum_{I=0}^N b_{I}\sum_{l=0,l\in even}^{2I}\mathcal{G}_{Il} \sum_{q,q'}\rho_{q}^{(l+3)/3}\rho_{q'}^{(2I-l+3)/3}\Bigg]\\
&\quad +\tilde{D}_0\Bigg[\sum_{I=0}^N b_{I}\sum_{l=0,l\in even}^{2I} \mathcal{G}_{Il} \sum_q \rho_{q}^{(2I+6)/3}\Bigg].
    \end{aligned}
\end{equation}

%\begin{equation}
%    \begin{aligned}
%u_{md}&=\tilde{C}_0\left(\frac{2}{(2\pi\hbar)^3}\right)^2(4\pi)^2\Bigg[\sum_{I=0}^N b_{I}\left(\hbar(3\pi^2)^{1/3}\right)^{2I+6}\\
%&\quad \sum_{k=0,k\in even}^{2I}\tbinom{2I}{k}\frac{1}{(k+3)}\frac{1}{(2I-k+3)}(\rho_n^{2I/3+2}\\
%&\quad +\rho_n^{k/3+1}\rho_p^{(2I-k+3)/3}+\rho_p^{k/3+1}\rho_n^{(2I-k+3)/3}+\rho_p^{2I/3+2})\Bigg]\\  
%&\quad +\tilde{D}_0\left(\frac{2}{(2\pi\hbar)^3}\right)^2(4\pi)^2 \Bigg[\sum_{I=0}^N b_{I}\left(\hbar(3\pi^2)^{1/3}\right)^{2I+6}\\
%&\quad  \sum_{k=0,k\in even}^{2I}\tbinom{2I}{k}\frac{1}{(k+3)}\frac{1}{(2I-k+3)}\\
%&\quad \left(\rho_n^{2I/3+2}+\rho_p^{2I/3+2}\right)\Bigg].
%C_0\left(\frac{2}{(2\pi\hbar)^3}\right)^2(4\pi)^2 \Bigg[\sum_{I=0}^4 b_{I}\sum_{k=0,k\in even}^{2I}\tbinom{2I}{k}\frac{1}{(k+3)}\\
%&\quad \frac{1}{(2I-k+3)}(p_{F_n}^{2I+6}+p_{F_n}^{k+3}p_{F_p}^{2I-k+3}+p_{F_p}^{k+3}p_{F_n}^{2I-k+3}+p_{F_p}^{2I+6})\Bigg]   \\
%&\quad +D_0\left(\frac{2}{(2\pi\hbar)^3}\right)^2(4\pi)^2 \Bigg[\sum_{I=0}^4 b_{I}\sum_{k=0,k\in even}^{2I}\tbinom{2I}{k}\\
%&\quad \frac{1}{(k+3)}\frac{1}{(2I-k+3)}\left(p_{F_n}^{2I+6}+p_{F_p}^{2I+6}\right)\Bigg]\\
%    \end{aligned}
%\end{equation}

By using the relation $\rho_q=\frac{\rho}{2}(1+\tau_q\delta)$, with $\tau_q=1 (-1)$ for neutron (proton), the above equation can be rewritten as follows, 
\begin{equation}
    \begin{aligned}
u_{md}&=\tilde{C}_0\Bigg[\sum_{I=0}^N b_{I}\sum_{l=0,l\in even}^{2I}\mathcal{G}_{Il}\times (\frac{\rho}{2})^{(2I+6)/3}  \\
&\quad \quad \quad \sum_{q,q'}(1+\tau_q\delta)^{(l+3)/3}(1+\tau_{q'}\delta)^{(2I-l+3)/3}\Bigg]\\
&+\tilde{D}_0\Bigg[\sum_{I=0}^N b_{I}\sum_{l=0,l\in even}^{2I} \mathcal{G}_{Il}\times (\frac{\rho}{2})^{(2I+6)/3}\\
&\quad \quad\quad \sum_q(1+\tau_q\delta)^{(2I+6)/3}\Bigg]\\
&\approx \tilde{C}_0\Bigg[\sum_{I=0}^N b_{I}\sum_{l=0,l\in even}^{2I}\mathcal{G}_{Il}\times (\frac{\rho}{2})^{(2I+6)/3} \\
&\quad \quad \quad  \sum_{q,q'}\left(1+\frac{l+3}{3}\tau_q\delta+\frac{1}{2}\frac{l+3}{3}\frac{l}{3}\delta^2\right)\\
&\left(1+\frac{(2I-l+3)}{3}\tau_{q'}\delta+\frac{1}{2}\frac{(2I-l+3)}{3}\frac{(2I-l)}{3}\delta^2\right)\Bigg]\\
&\quad +\tilde{D}_0\Bigg[\sum_{I=0}^N b_{I}\sum_{l=0,l\in even}^{2I} \mathcal{G}_{Il}\times(\frac{\rho}{2})^{(2I+6)/3} \\
&\quad\quad  \sum_q (1+\frac{2I+6}{3}\tau_q\delta+\frac{1}{2}\frac{2I+6}{3}\frac{2I+3}{3}\delta^2)\Bigg]\\
%&=\tilde{C}_0\Bigg[\sum_{I=0}^N b_{I}\sum_{k=0,k\in even}^{2I}\mathcal{G}_{Ik} \\
%& (\frac{\rho}{2})^{(2I+6)/3} \sum_{q,q'}\left(1+\frac{k+3}{3}\tau_q\delta+\frac{k+3}{3}\frac{k}{3}\delta^2\right)\\
%&\left(1+\frac{(2I-k+3)}{3}\tau_{q'}\delta+\frac{(2I-k+3)}{3}\frac{(2I-k)}{3}\delta^2\right)\Bigg]\\
%&+\tilde{D}_0\Bigg[\sum_{I=0}^N b_{I}\sum_{k=0,k\in even}^{2I} \mathcal{G}_{Ik} \\
%&\quad \quad \times (\frac{\rho}{2})^{(2I+6)/3}\sum_q (1+\frac{2I+6}{3}\tau_q\delta+\frac{2I+6}{3}\frac{2I+3}{3}\delta^2)\Bigg].\\
    \end{aligned}
\end{equation}

With the parabolic approximation, the $u_{md}$ can be rewritten as
\begin{equation}
    u_{md}= u_{md}^0+u_{md}^{asy}\delta^2+\cdots,
\end{equation}

Thus, the $u_{md}^0$ is,
\begin{equation}\label{umd}
    \begin{aligned}
&u_{md}^0=4(\tilde{C}_0+\frac{\tilde{D}_0}{2})\Bigg[\sum_{I=0}^N b_{I}\sum_{l=0,l\in even}^{2I}\mathcal{G}_{Il}\times(\frac{\rho}{2})^{(2I+6)/3}\Bigg].
%&u_{md}^0=\tilde{C}_0\left(\frac{2}{(2\pi\hbar)^3}\right)^2(4\pi)^2 \Bigg[\sum_{I=0}^N b_{I}\left(\hbar(3\pi^2)^{1/3}\right)^{2I+6}\\
%&\quad \times \sum_{k=0,k\in even}^{2I}\tbinom{2I}{k}\frac{1}{(k+3)}\frac{1}{(2I-k+3)}(\frac{\rho}{2})^{2I/3+2}\times 4\Bigg]\\  
%&\quad \quad +\tilde{D}_0\left(\frac{2}{(2\pi\hbar)^3}\right)^2(4\pi)^2 \Bigg[\sum_{I=0}^N b_{I}\left(\hbar(3\pi^2)^{1/3}\right)^{2I+6} \\
%&\quad \times \sum_{k=0,k\in even}^{2I}\tbinom{2I}{k}
%\frac{1}{(k+3)}\frac{1}{(2I-k+3)}(\frac{\rho}{2})^{2I/3+2}\times 2\Bigg]\\
%&=4(\tilde{C}_0+\frac{\tilde{D}_0}{2})\left(\frac{2}{(2\pi\hbar)^3}\right)^2(4\pi)^2 \Bigg[\sum_{I=0}^N b_{I}\left(\hbar(3\pi^2)^{1/3}\right)^{2I+6}\\
%&\quad \times \sum_{k=0,k\in even}^{2I}\tbinom{2I}{k}\frac{1}{(k+3)}\frac{1}{(2I-k+3)}(\frac{\rho}{2})^{(2I+6)/3}\Bigg]\\
%&=4(\tilde{C}_0+\frac{\tilde{D}_0}{2})\Bigg[\sum_{I=0}^N b_{I}\sum_{k=0,k\in even}^{2I}\mathcal{G}_{Ik}\times(\frac{\rho}{2})^{(2I+6)/3}\Bigg].
    \end{aligned}
\end{equation}
The asymmetry term $u_{md}^{asy}$ is,
\begin{equation}\label{umd_asy}
    \begin{aligned}
&u_{md}^{asy}=2\tilde{C}_0\Bigg[\sum_{I=0}^N b_{I}\sum_{l=0,l\in even}^{2I}\mathcal{G}_{Il}\times(\frac{\rho}{2})^{(2I+6)/3}\\
&\quad \quad \quad\quad\quad \Big(\frac{l+3}{3}\frac{l}{3}+\frac{2I-l+3}{3}\frac{2I-l}{3}\Big)\Bigg]\\
&\quad +\tilde{D}_0\Bigg[\sum_{I=0}^N b_{I}\frac{2I+6}{3}\frac{2I+3}{3}\sum_{l=0,l\in even}^{2I} \mathcal{G}_{Il}\times(\frac{\rho}{2})^{(2I+6)/3} \Bigg]\\
%\tilde{C}_0\left(\frac{2}{(2\pi\hbar)^3}\right)^2(4\pi)^2\\
%&\quad \quad \Bigg[\sum_{I=0}^N b_{I}\left(\hbar(\frac{3\pi^2}{2})^{1/3}\right)^{2I+6}\sum_{k=0,k\in even}^{2I}\tbinom{2I}{k}\\
%&\quad \quad 2\times\Big(\frac{2I-k+3}{3}\frac{2I-k}{3}+\frac{k+3}{3}\frac{k}{3}\Big)\\
%&\quad \quad \quad \times \frac{1}{(k+3)}\frac{1}{(2I-k+3)}\rho^{2I/3+2}\Bigg].\\
%&\quad \quad+\tilde{D}_0\left(\frac{2}{(2\pi\hbar)^3}\right)^2(4\pi)^2\Bigg[\sum_{I=0}^N b_{I}\left(\hbar(\frac{3\pi^2}{2})^{1/3}\right)^{2I+6}\\
%&\quad \quad (\frac{2I}{3}+2)(\frac{2I}{3}+1) \sum_{k=0,k\in even}^{2I}\tbinom{2I}{k}
%\frac{1}{(k+3)}\frac{1}{(2I-k+3)}\\
%&\quad \quad \rho^{2I/3+2}\Bigg].
    \end{aligned}
\end{equation}

\section{single-particle potential}\label{appendix:Vq}
The non-local part of single-particle potential of nucleon in cold uniform nuclear matter is:
\begin{eqnarray}
     \begin{aligned}
    V^\text{md}_q & (\rho,\delta,p) = \frac{\delta u_\text{md}}{\delta
        f_q} \\
    &= 2\tilde{C}_0\int d^{3}p^{'}f(\mathbf{r},\mathbf{p'})g(\mathbf{p}-\mathbf{p'})\\
    &\quad+2\tilde{D}_{0}\int_{\mathbf{p'}<p_{F_q}} d^{3}p^{'}f_{q}(\mathbf{r},\mathbf{p'})g(\mathbf{p}-\mathbf{p'})\\
    &= 2\tilde{C}_0\sum_{\tau=n,p}\int_{\mathbf{p'}<p_{F_\tau}} d^{3}p^{'}f_\tau(\mathbf{r},\mathbf{p'})g(\mathbf{p}-\mathbf{p'})\\
    &\quad +2\tilde{D}_{0}\int_{\mathbf{p'}<p_{F_q}} d^{3}p^{'}f_{q}(\mathbf{r},\mathbf{p'})g(\mathbf{p}-\mathbf{p'})\\ 
    \end{aligned}
\end{eqnarray}

For the $\tilde{C}_0$ term,
\begin{equation}
    \begin{aligned}
&2\tilde{C}_0\int \text{d}^3p' f(\mathbf{r},\mathbf{p'})g(\mathbf{p}-\mathbf{p'}) \\
%&=2C_0\left(\frac{4}{(2\pi\hbar)^3}\right)4\pi \int_{0}^{p_F}p'^2\text{d}p' \\
%&\quad\quad\big(\sum_{I=0}^4 b_{I}\sum_{k=0}^{2I}\tbinom{2I}{k}\mathbf{p}^k\cdot(-\mathbf{p'})^{2I-k}\big).
&=2\tilde{C}_0\left(\frac{2}{(2\pi\hbar)^3}\right)4\pi \Bigg[\sum_{I=0}^N b_{I}\sum_{l=0,l\in even}^{2I}\tbinom{2I}{l}\\
&\quad \quad \frac{1}{(2I-l+3)}(p_{F_n}^{2I-l+3}+p_{F_p}^{2I-l+3})\times p^l\Bigg]\\
&=2\tilde{C}_0\Bigg[\sum_{I=0}^N b_{I}\sum_{l=0,l\in even}^{2I}\mathcal{\Tilde{A}}_{Il}\sum_q\rho_q^{(2I-l+3)/3}\times p^l\Bigg]
    \end{aligned}
\end{equation}
here, 
\begin{equation*}
\begin{aligned}
\mathcal{\Tilde{A}}_{Il}&=\frac{2}{(2\pi\hbar)^3}4\pi\Bigg[\left(\hbar(3\pi^2)^{1/3}\right)^{(2I-l+3)} \tbinom{2I}{l}\frac{1}{(2I-l+3)}\Bigg], 
\end{aligned}
\end{equation*}
Similarly, the $\tilde{D}_0$ terms,
\begin{equation}
    \begin{aligned}
&2\tilde{D}_0\int \text{d}^3p' f_q(\mathbf{r},\mathbf{p'})g(\mathbf{p}-\mathbf{p'}) \\
%&=2D_0\left(\frac{2}{(2\pi\hbar)^3}\right)4\pi \int_{0}^{p_{F_q}}p'^2\text{d}p' \\
%&\quad\quad\big(\sum_{I=0}^4 b_{I}\sum_{k=0}^{2I}\tbinom{2I}{k}\mathbf{p}^k\cdot(-\mathbf{p'})^{2I-k}\big)\\
&=2\tilde{D}_0\left(\frac{2}{(2\pi\hbar)^3}\right)4\pi \Bigg[\sum_{I=0}^N b_{I}\sum_{l=0,l\in even}^{2I}\tbinom{2I}{l}\\
&\quad \quad \frac{1}{(2I-l+3)}p_{F_q}^{2I-l+3}\times p^l\Bigg]\\
&=2\tilde{D}_0\Bigg[\sum_{I=0}^N b_{I}\sum_{l=0,l\in even}^{2I}\mathcal{\Tilde{A}}_{Il}\times\rho_q^{(2I-l+3)/3}\times p^l\Bigg]
    \end{aligned}
\end{equation}

The $V^\text{md}_q$ will be,
\begin{equation}
    \begin{aligned}
V^\text{md}_q(\rho,\delta,p)&=2\tilde{C}_0\Bigg[\sum_{I=0}^N b_{I}\sum_{l=0,l\in even}^{2I}\mathcal{\Tilde{A}}_{Il}\sum_q\rho_q^{(2I-l+3)/3}p^l\Bigg]\\
&\quad+2\tilde{D}_0\Bigg[\sum_{I=0}^N b_{I}\sum_{l=0,l\in even}^{2I}\mathcal{\Tilde{A}}_{Il}\rho_q^{(2I-l+3)/3}p^l\Bigg]
%2\tilde{C}_0\left(\frac{2}{(2\pi\hbar)^3}\right)4\pi \Bigg[\sum_{I=0}^N b_{I}\sum_{k=0,k\in even}^{2I}\tbinom{2I}{k}\\
%&\quad \quad \frac{1}{(2I-k+3)}(p_{F_n}^{2I-k+3}+p_{F_p}^{2I-k+3})\times p^k\Bigg] \\
%&\quad \quad +2\tilde{D}_0\left(\frac{2}{(2\pi\hbar)^3}\right)4\pi \Bigg[\sum_{I=0}^4 b_{I}\sum_{k=0,k\in even}^{2I}\tbinom{2I}{k}\\
%&\quad \quad \frac{1}{(2I-k+3)}p_{F_q}^{2I-k+3}\times p^k\Bigg]\\
%&=2\tilde{C}_0\left(\frac{2}{(2\pi\hbar)^3}\right)4\pi\Bigg[\sum_{I=0}^N b_{I}\left(\hbar(3\pi^2)^{1/3}\right)^{2I-k+3}\\
%&\quad  \sum_{k=0,k\in even}^{2I}\tbinom{2I}{k}(\rho_n^{(2I-k+3)/3}+\rho_p^{(2I-k+3)/3})\\
%&\quad \frac{1}{(2I-k+3)}\times p^k\Bigg]\\  
%&\quad  +2\tilde{D}_0\left(\frac{2}{(2\pi\hbar)^3}\right)4\pi \Bigg[\sum_{I=0}^N b_{I}\left(\hbar(3\pi^2)^{1/3}\right)^{2I-k+3}\\
%&\quad \sum_{k=0,k\in even}^{2I}\tbinom{2I}{k}
%\frac{1}{(2I-k+3)}\rho_q^{(2I-k+3)/3}\times p^k.
    \end{aligned}
\end{equation}

By using the relation $\rho_q=\frac{\rho}{2}(1+\tau_q\delta)$, with $\tau_q=1$ or $-1$ for neutron and protons, the above equation can be rewritten as follows,

\begin{equation}
    \begin{aligned}
V^\text{md}_q(\rho,\delta,p)&=2\tilde{C}_0\Bigg[\sum_{I=0}^N b_{I}\sum_{l=0,l\in even}^{2I}\mathcal{\Tilde{A}}_{Il}\times(\frac{\rho}{2})^{(2I-l+3)/3}\\
&\quad \sum_q(1+\tau_q\delta)^{(2I-l+3)/3}\times p^l\Bigg]\\
&\quad  +2\tilde{D}_0\Bigg[\sum_{I=0}^N b_{I}\sum_{l=0,l\in even}^{2I}\mathcal{\Tilde{A}}_{Il}\times(\frac{\rho}{2})^{(2I-l+3)/3}\\
&\quad   (1+\tau_q\delta)^{(2I-l+3)/3}\times p^l\Bigg]\\
&\approx2\tilde{C}_0\Bigg[\sum_{I=0}^N b_{I}\sum_{l=0,l\in even}^{2I}\mathcal{\Tilde{A}}_{Il}\times(\frac{\rho}{2})^{(2I-l+3)/3}\\
&\quad  \sum_q(1+\frac{2I-l+3}{3}\tau_q\delta+\cdots)\times p^l\Bigg]\\
& \quad   +2\tilde{D}_0\Bigg[\sum_{I=0}^N b_{I}\sum_{l=0,l\in even}^{2I}\mathcal{\Tilde{A}}_{Il}\times(\frac{\rho}{2})^{(2I-l+3)/3}\\
&\quad (1+\frac{2I-l+3}{3}\tau_q\delta+\cdots)\times p^l\Bigg]\\
    \end{aligned}
\end{equation}

The $V_{md}$ can be rewritten as
\begin{equation}
    V_{md}= V_{md}^0\pm V_{md}^{sym}\delta+\cdots.
\end{equation}

Thus, the $V_{md}^0$ is,
\begin{equation}\label{V_{md}^0}
    \begin{aligned}
   V_{md}^0&=4(\tilde{C}_0+\frac{\tilde{D}_0}{2})\\
   &\quad \quad \Bigg[\sum_{I=0}^N b_{I}\sum_{l=0,l\in even}^{2I}\mathcal{\Tilde{A}}_{Il}\times(\frac{\rho}{2})^{(2I-l+3)/3}\times p^l\Bigg]\\
   \end{aligned}
\end{equation}

The asymmetry term $V_{md}^{sym}$ is
\begin{equation}\label{V_{md}^{sym}}
    \begin{aligned}
   V_{md}^{sym}&=2\tilde{D}_0\Bigg[\sum_{I=0}^N b_{I}\sum_{l=0,l\in even}^{2I}\mathcal{\Tilde{A}}_{Il}\times(\frac{\rho}{2})^{(2I-l+3)/3}\\
&\quad \frac{2I-l+3}{3}\times p^l\Bigg]\\
   \end{aligned}
\end{equation}

\section{Relation between the effective mass splitting and symmetry potential}
\label{dmnp-vasy}

According to Eq.(\ref{effm}), 
\begin{equation}
\begin{aligned}
\label{eq:effm-splitting}
    \frac{m}{m_n^*}-\frac{m}{m_p^*}&=2m\big(\frac{\partial V_n}{\partial p^2}-\frac{\partial V_p}{\partial p^2}\big)=4m\delta \frac{\partial V_{sym}}{\partial p^2}.
\end{aligned}    
\end{equation}

In addition, 
\begin{equation}
    \begin{aligned}
      \frac{m}{m_n^*}-\frac{m}{m_p^*}=-\frac{m(m_n^*-m_p^*)}{m_n^*m_p^*}\approx-(\frac{m}{m^*})^2\Delta m_{np}^*.
    \end{aligned}
\end{equation}
The approximation comes from the $\frac{m_q^*}{m}=\frac{m^*}{m}\pm\frac{\delta m^*}{m}$, and the product of $\frac{m_n^*m_p^*}{m^2}=(\frac{m^*}{m})^2-(\frac{\delta m^*}{m})^2\approx (\frac{m^*}{m})^2$.

\section{High order expansion coefficients of nuclear matter equation of state}\label{EOS-Taylor}

\begin{table*}[htbp]
\centering
\caption{The parameters $K_{0}$, $Q_{0}$, $Z_{0}$, $K_{sym}$, $Q_{sym}$, $Z_{sym}$ are in MeV.}
\label{tab:nmpara-Tayl}
\begin{tabular}{cccccc} \\ 
 \hline
 \hline
 Para.& (0.3, 46) & (-0.3,46) & (0.3, 100) & (-0.3, 100)  \\ 
 \hline
 $K_{0}$ &\multicolumn{4}{c}{230 (230)}&   \\ 
 $Q_{0}$ &\multicolumn{4}{c}{-406.07 (-376.62)}&    \\ 
 $Z_{0}$ &\multicolumn{4}{c}{1651.85 (1629.82)}& \\
 $K_{sym}$ & -162.21 (-122.96) &-174.23 (-182.36)  &41.72 (74.56) & 29.70 (15.16) \\ 
 $Q_{sym}$ & 363.94 (526.53) &422.23 (368.67) &-89.49 (63.90) &-31.20 (-93.97)   \\ 
 $Z_{sym}$ & -2403.24(-3173.67) &-2203.28(-2033.33)  &-34.78 (-702.16) & 165.19(438.18)  \\ \hline
\end{tabular}%
\end{table*}

Around the nuclear matter saturation density $\rho_0$, the binding energy per nucleon in symmetric nuclear matter can be expanded (e.g., up to fourth-order in density) as:
\begin{equation}\label{eq:EOS-Tayl}
    E/A(\rho)=E_0(\rho_0)+\frac{K_0}{2!}x^2+\frac{Q_0}{3!}x^3+\frac{Z_0}{4!}x^4+O(x^5),  
\end{equation}
where $x$ is a dimensionless variable, $x=\frac{\rho-\rho_0}{3\rho_0}$.

The incompressibility $K_0$ is,
\begin{equation}\label{K0-Tayl}
\begin{aligned}
K_0&=9\rho_0^2\frac{\partial^2 E/A}{\partial \rho^2}|_{\rho_0}\\
&=\frac{-6}{5}\epsilon_{F}^0+9\frac{\beta}{\gamma+1}\gamma(\gamma-1)\\
&\quad+6(\tilde{C}_0+\frac{\tilde{D}_0}{2})\sum_{I=1}^N\tilde{g}_{md}^{(I)}\left(\frac{2I}{3}+1\right)I\rho_0^{2I/3+1}.
\end{aligned}
\end{equation}

The skewness coefficient $Q_0$ is,
\begin{equation}\label{Q0-Tayl}
\begin{aligned}
Q_0&=27\rho_0^3\frac{\partial^3 E/A}{\partial \rho^3}|_{\rho_0}\\
&=\frac{24}{5}\epsilon_{F}^0+27\frac{\beta}{\gamma+1}\gamma(\gamma-1)(\gamma-2)\\
&\quad+27(\tilde{C}_0+\frac{\tilde{D}_0}{2})\sum_{I=1}^N\tilde{g}_{md}^{(I)}\left(\frac{2I}{3}+1\right)\frac{2I}{3}\frac{2I-3}{3}\rho_0^{2I/3+1}.
\end{aligned}
\end{equation}

The fourth derivative of the energy per nucleon $Z_0$ is,
\begin{equation}\label{Z0-Tayl}
\begin{aligned}
Z_0&=81\rho_0^4\frac{\partial^4 E/A}{\partial \rho^4}|_{\rho_0}\\
&=\frac{-168}{5}\epsilon_{F}^0+81\frac{\beta}{\gamma+1}\gamma(\gamma-1)(\gamma-2)(\gamma-3)\\
&\quad+81(\tilde{C}_0+\frac{\tilde{D}_0}{2})\sum_{I=1}^N\tilde{g}_{md}^{(I)}\left(\frac{2I}{3}+1\right)\\
&\quad\times\frac{2I}{3}\frac{2I-3}{3}\frac{2I-6}{3}\rho_0^{2I/3+1}.
\end{aligned}
\end{equation}

Around the normal nuclear density $\rho_0$, the nuclear symmetry energy S($\rho$) can be similarly expanded (e.g., up to fourth order in $x$) as:
\begin{equation}\label{eq:Srho-Tayl}
    S(\rho)=S(\rho_0)+Lx+\frac{K_{sym}}{2!}x^2+\frac{Q_{sym}}{3!}x^3+\frac{Z_{sym}}{4!}x^4+O(x^5).   
\end{equation}

The slope of the symmetry energy $L$ is
\begin{equation}\label{L-Tayl}
\begin{aligned}
    L&=3\rho_0\frac{\partial S(\rho)}{\partial \rho}|_{\rho_0}\\
    &=\frac{2}{3}\epsilon_{F}^0+3A_{sym}+3B_{sym}\gamma\\
    &\quad +3\sum_{I=1}^N\tilde{C}_{sym}^{(I)}\left(\frac{2I}{3}+1\right)\rho_0^{2I/3+1}.
\end{aligned}
\end{equation}

The curvature of the symmetry energy $K_{sym}$ is
\begin{equation}\label{Ksym-Tayl}
\begin{aligned}
K_{sym}&=9\rho_0^2\frac{\partial^2 S(\rho)}{\partial \rho^2}|_{\rho_0}\\
&=-\frac{2}{3}\epsilon_{F}^0+9B_{sym}\gamma(\gamma-1)\\
&\quad +9\sum_{I=1}^N\tilde{C}_{sym}^{(I)}\left(\frac{2I}{3}+1\right)\frac{2I}{3}\rho_0^{2I/3+1}.
\end{aligned}
\end{equation}

The third derivative of the symmetry energy $Q_{sym}$ is
\begin{equation}\label{Qsym-Tayl}
\begin{aligned}
Q_{sym}&=27\rho_0^3\frac{\partial^3 S(\rho)}{\partial \rho^3}|_{\rho_0}\\
&=\frac{8}{3}\epsilon_{F}^0+27B_{sym}\gamma(\gamma-1)(\gamma-2)\\
&\quad +27\sum_{I=1}^N\tilde{C}_{sym}^{(I)}\left(\frac{2I}{3}+1\right)\frac{2I}{3}\frac{2I-3}{3}\rho_0^{2I/3+1}.
\end{aligned}
\end{equation}

The fourth derivative of the symmetry energy $Z_{sym}$ is
\begin{equation}\label{Zsym-Tayl}
\begin{aligned}
Z_{sym}&=81\rho_0^4\frac{\partial^4 S(\rho)}{\partial \rho^4}|_{\rho_0}\\
&=\frac{-56}{3}\epsilon_{F}^0+81B_{sym}\gamma(\gamma-1)(\gamma-2)(\gamma-3)\\
&\quad +81\sum_{I=1}^N\tilde{C}_{sym}^{(I)}\left(\frac{2I}{3}+1\right)\frac{2I}{3}\frac{2I-3}{3}\frac{2I-6}{3}\rho_0^{2I/3+1}.
\end{aligned}
\end{equation}

In Table\ref{tab:nmpara-Tayl}, we list the values of these higher-order expansion coefficients of the nuclear matter equation that we used in this work. The values in brackets from the second to the seventh rows represent the values obtained with the standard Skyrme MDI.

%\section*{}
%\begin{thebibliography}{100}
%\end{thebibliography}
\bibliography{References}

\end{document}